\documentclass[aps,prl,twocolumn,superscriptaddress,showpacs,floatfix]{revtex4-2}
\usepackage{amsmath,amssymb,amsfonts,dsfont, float,graphics,epsfig,epstopdf,color,verbatim,tabularx,bm,multirow}
\usepackage{wasysym}
\usepackage[utf8]{inputenc}
\usepackage[T1]{fontenc}
\usepackage{xcolor}
\usepackage{subfigure}
\usepackage{graphicx}
\usepackage{young}
\usepackage{youngtab}
\usepackage{ytableau}
\usepackage[normalem]{ulem}

\IfFileExists{newtxtext.sty}
   {\usepackage{newtxtext,newtxmath}}
   {\IfFileExists{stix.sty}
      {\usepackage{stix}}
      {\IfFileExists{mathptmx.sty}
      {\usepackage{mathptmx}}{} } }

\usepackage{hyperref}
\hypersetup{
	colorlinks = true,
	linkcolor = [rgb]{0.70,0.13,0.13},
	citecolor = [rgb]{0.13,0.55,0.13},
	urlcolor = [rgb]{0.25,0.41,0.88}}

\DeclareSymbolFont{sfletters}{OML}{cmbrm}{m}{it}
\DeclareMathSymbol{\sfeps}{\mathord}{sfletters}{"22}
\newcommand{\Tr}{\operatorname{Tr}}

\graphicspath{{Figures/}}

\begin{document}
\title{Detecting the (emergent) continuous symmetry of criticality via a subsystem's entanglement spectrum}

\author{Bin-Bin Mao}
\thanks{These authors contributed equally to this work.}
\affiliation{Department of Physics, School of Science and Research Center for Industries of the Future, Westlake University, Hangzhou 310030, China}
\affiliation{School of Foundational Education, University of Health and Rehabilitation Sciences, Qingdao 266000, China}

\author{Zhe Wang}
\thanks{These authors contributed equally to this work.}
\affiliation{Department of Physics, School of Science and Research Center for Industries of the Future, Westlake University, Hangzhou 310030, China}
\affiliation{Institute of Natural Sciences, Westlake Institute for Advanced Study, Hangzhou 310024, China}

\author{Bin-Bin Chen}
\email{bchen@buaa.edu.cn}
\affiliation{Peng Huanwu Collaborative Center for Research and Education, Beihang University, Beijing 100191, China}

\author{Zheng Yan}
\email{zhengyan@westlake.edu.cn}
\affiliation{Department of Physics, School of Science and Research Center for Industries of the Future, Westlake University, Hangzhou 310030, China}
\affiliation{Institute of Natural Sciences, Westlake Institute for Advanced Study, Hangzhou 310024, China}

\begin{abstract}
The (emergent) symmetry of a critical point constitutes fundamental pieces of information for determining the universality class and effective field theory. 
However, the underlying symmetry thus far can be conjectured only indirectly from the dimension of the order parameters in symmetry-breaking phases, 
and its correctness requires further verification to avoid overlooking hidden order parameters, which by itself is also a difficult task.
In this work, we introduce an unbiased numerical approach to identify the underlying (emergent) symmetry of a critical point in quantum many-body systems without prior knowledge about the corresponding low-energy effective field theory. 
By numerically calculating the reduced density matrix in a very small subsystem of the total system, the Anderson tower of states in the entanglement spectrum can be obtained, clearly reflecting the underlying (emergent) symmetry of criticality. 
This is attributed to the fact that the entanglement spectrum can reveal the broken symmetry of the ground state of entanglement Hamiltonian after cooling from the critical point along an extra entanglement-temperature axis.
\end{abstract}
\date{\today}
\maketitle

{\it{\color{blue}Introduction.-}}
Phase transitions, spontaneous symmetry breaking (SSB) and the universality class of criticality are the core concepts in condensed matter and statistical physics~\cite{sachdev1999quantum,wannier1987statistical,girvin2019modern}. For an unknown phase transition, we need to comprehensively consider its symmetry, dimension and other information to analyze its possible universality class of phase transitions. However, identifying the universality of a phase transition is challenging since the low-energy effective theory of a system can have a different symmetry to the Hamiltonian. Higher symmetry may emerge at various phase transition, including the first-order phase transition point~\cite{BWZhao2019,yan2021widely,Sun_2021}, continuous phase transition point~\cite{ding2016emergent}, multicritical point~\cite{Zhou_2025} or deconfined quantum criticality (DQC)~\cite{senthilDeconfined2004, senthilQuantum2004,sandvikEvidence2007,nahumDeconfined2015,qinDuality2017,wangDeconfined2017,senthilDeconfined2023}.

For example, the symmetries associated with phases proximate to the deconfined quantum critical point (DQCP) are incompatible with the Landau-Ginzburg-Wilson (LGW) framework of phase transition. Such DQCP-scenario phase transitions, by definition, are continuous and associated with the emergence of higher symmetries~\cite{nahumEmergent2015,tanaka2005many,senthil2006competing,liu2024emergent,huang2019emergent,wang2017deconfined,ma2019role,senthil2024deconfined,d2024entanglement}. Thus, the DQCP provides a novel paradigm beyond LGW phase transition theory, which has stimulated significant theoretical advances, including a connection to higher-dimensional symmetry-protected topological states, the 't Hooft anomaly~\cite{ashvinsenthil}, and fractionalized degrees of freedom~\cite{senthilfisher,nahumEmergent2015,maRole2019,sreejithEmergent2019,maDynamics2018}. Extensive numerical ~\cite{haradaPossibility2013, sandvikEvidence2007, louvbsneel2009, liuSuperconductivity2019, liaoDiracI2022, shaoQuantum2016, maDynamics2018,zhang2018continuous,liu2024deconfined,liu2022emergence} and experimental ~\cite{jimenezquantum2021, zayed4spin2017, guoQuantum2020, sunEmergent2021, cuiProximate2023,guoDeconfined2023,cui2025deconfined} progress concerning the topic of DQCP has been reported during the past decade.

Despite the extensive debate regarding the DQCP mechanism~\cite{takahashi2024so,zhaoScaling2022,deng2024diagnosing,song2023deconfined,zhou2024mathrmso5}, our focus in the present work lies in the determination of the intrinsic symmetries at phase transition points. For instance, by analyzing the order parameter dimensionality in the N\'eel phase and valence bond solid (VBS) phase of (2+1)D quantum systems combined with field theory analysis, an emergent $SO(5)$ symmetry can be theoretically conjectured and numerically verified~\cite{nahumEmergent2015}. However, a {\it prior} assumption of an order parameter may overlook some hidden dimensions; e.g., when a system possesses $O(4)$ symmetry, we can still obtain an $O(3)$-symmetric order-parameter histogram even though we incorrectly assume only a three-dimensional order parameter, which will then lead us to an incorrect conclusion of an $O(3)$-symmetric system.
Here, we aim to establish an unbiased approach to probe the underlying (emergent) symmetry at a critical point---without prior knowledge of its low-energy effective field theory---even in large-scale or/and high-dimensional quantum many-body systems.

A desirable approach would involve spontaneous continuous symmetry breaking, whereby the Anderson tower of state (TOS) {even in small finite systems} would then be used to identify the type of broken continuous symmetry~\cite{anderson1952approximate,lhuillier2005frustratedquantummagnets,wietek2017studying}.
However, we find that this approach would be impractical for a quantum phase transition with emergent symmetry, e.g., at a DQCP.
On the one hand, when we directly probe the critical point, although the emergent symmetry is preserved, the low-lying spectrum is dominated by critical modes, such as a conformal field theory (CFT) tower instead of an Anderson tower~\footnote{A CFT tower is formed if conformal field theory is valid for critical points; for the (2+1)D $J$-$Q$ model of a DQCP, this remains controversial}.
On the other hand, by tuning the Hamiltonian parameters to shift the system away from criticality and to potentially achieve SSB, we may concurrently hinder the emergence of higher symmetry.

The entanglement spectrum (ES) provides a solution. Similar to the energy spectrum, the ES is also expected to exhibit an Anderson TOS when continuous symmetry breaking occurs in the ground state~\cite{Alba2013entanglement,Kolley2013entanglement,mao2025sampling,rams2018precise}.
In this work, we will show that, in addition to ground states with broken symmetry, the ES displays a TOS even at a phase transition point, and the TOS reflects the (emergent) symmetry of the transition point. 
The reason lies in the fact that an entanglement Hamiltonian (EH) naturally introduces an effective temperature $T_E$~\cite{chandran2013how,Poilblanc2010entanglement,zyan2021entanglement,song2023different,li2023relevant,wu2023classical} and a thermal phase transition at $T_E=1$ towards a (emergent) symmetry-breaking state. 
The ES can then manifest such symmetry-breaking physics through the low-lying ground-state manifold of the EH (see the discussion below).
%The ES actually describes the zero-temperature physics of the EH and can be treated as cooling from the phase transition point and spontaneously breaking the symmetry there.

{\it{\color{blue} A general phase diagram of entanglement Hamiltonian.}} A quantum system can be bipartionally separated into a subsystem $A$ and an environment $B$. Therefore, we can define the reduced density matrix (RDM) as $\rho_A=\Tr_B(\rho)$, where $\rho$ is the density matrix of the total system. The EH was further defined as $H_E=-\ln(\rho_A)$, and its eigenvalue spectrum is ES~\cite{Li2008entangle,Poilblanc2010entanglement,XLQi2012}. According to the definitions of EH and ES, $\rho_A=e^{-H_E}$ can be treated as a Gibbs mixed state of EH at $T_E=1$, where $T_E$ is the effective temperature for EH~\cite{zyan2021entanglement,song2023different}. 
Thus, the phase diagram of EH at a fixed effective temperature $T_E=1$ is the same as the ground-state phase diagram of the original Hamiltonian (OH).

%\red{Strictly, it is not so general, for example, the phase transition line can also be lower than $T_E=1$. The generality here is that there must be a spontaneous-symmetry-breaking phase at $T_E=0$ which can reflect the symmetry of the finite $T_E$ phase transition point.}

\begin{figure}[htp]
\centering
\includegraphics[width=0.48\textwidth]{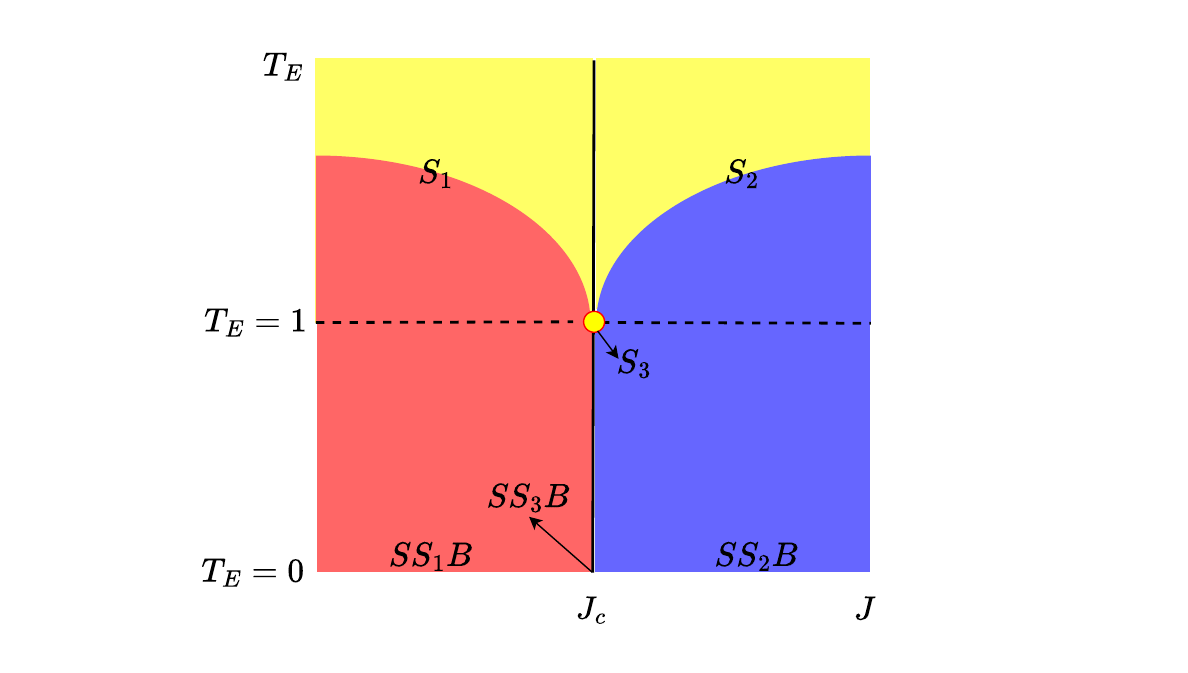}
\caption{A general phase diagram of the EH. $J$ is a tunable parameter of the original Hamiltonian (OH). $T_E$ is the temperature of the EH. The reduced density matrix $\rho_A=e^{-H_E}$ can be considered a Gibbs mixed state of EH at $T_E=1$, and the dashed line $T_E=1$ is the ground-state phase of the OH. The yellow point is not only the quantum phase transition point of the OH but also the thermal phase transition point of the EH. In this way, the EH naturally introduces an extra temperature space, and SSB will occur through reducing $T_E$ (cooling) from the upper critical curves or points. Because the ES reflects SSB in the ground state at $T_E=0$, different symmetries ($S_1$, $S_2$, $S_3$) in finite $T_E$ phase transitions can be detected.}
\label{fig1}
\end{figure}

Furthermore, a comprehensive finite-$T_E$ phase diagram of the EH can be deduced~\footnote{Noting that the phase transition line can also be lower than $T_E=1$ if the phase at $T_E=1$ is disordered.}, as shown in Fig.~\ref{fig1}. {There is a spontaneous-symmetry-breaking phase at $T_E=0$ which can reflect the symmetry of the finite $T_E$ phase transition point.}
The horizontal $J$ axis is a controlling parameter in the OH driving a quantum phase transition between two symmetry-breaking phases $SS_1B$ and $SS_2B$ with different symmetries $S_1$ and $S_2$. As stated above, the dashed line $T_E=1$ in the $T_E-J$ phase diagram of the EH also manifests the ground-state phase diagram of the OH, where the yellow dot at $J=J_c$ and $T_E=1$ reflects the quantum critical point of the OH, 
and an emergent symmetry $S_3$ is required by the DQCP scenario.
Due to the divergent correlation length at the yellow dot, this suggests a thermal phase transition along the vertical effecitve temperature axis $T_E$ towards a spontaneous symmetry-breaking phase $SS_3B$. 

Henceforth, the ground-state phase diagram of the EH will be three symmetry-breaking phases ($SS_1B$, $SS_2B$, $SS_3B$) as denoted in the $T_E=0$ line of Fig.~\ref{fig1}, which can potentially detected by the ES. 
Nevertheless, whether and how the $SS_3B$ induced by such an effecitve cooling process can reflect the emergent symmetry $S_3$ at the thermal phase transition point (i.e., the yellow dot in Fig.~\ref{fig1}) remains an open question. 
We will then answer this question and demonstrate the SSB phenomenon for emergent symmetry through our detailed numerical results below.

\begin{figure}[htp]
\centering
\includegraphics[width=0.48\textwidth]{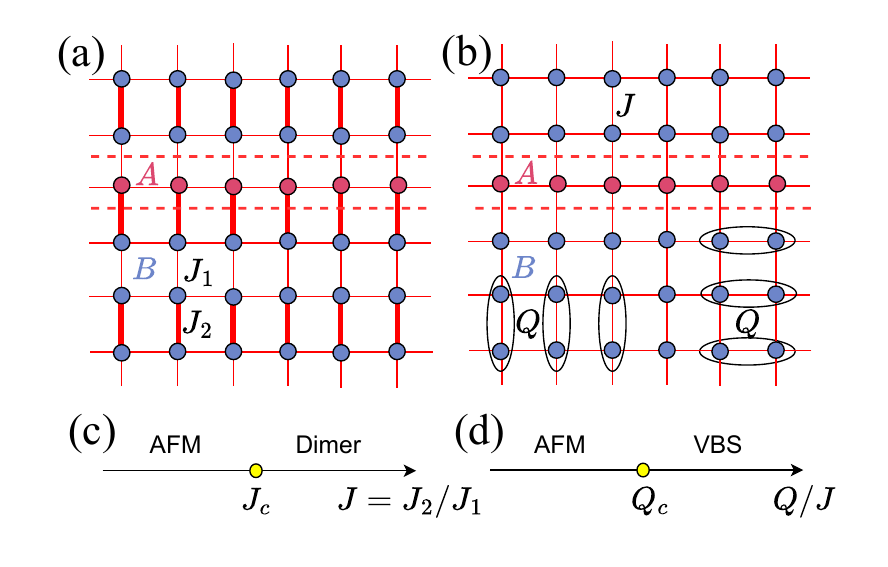}
\caption{Model and phase diagram. (a) Spin-1/2 dimerized AFM Heisenberg model: the strong bonds $J_2>0$ are indicated by thick lines, and the weak bonds $J_1> 0$ are indicated by thin lines. The dashed lines are used to illustrate the bipartition into two subsystems. The red dashed lines illustrate the cutting method, in which $A$ is a ring and the other part is denoted as $B$. (b) $J-Q$ model: $J>0$ is the AFM Heisenberg interaction, and the six-spin $Q$ interaction covers the entire lattice. (c) Diagram of the dimerized AFM Heisenberg model with $J_1=1$, in which the quantum critical point is $ J_c=J_{2}/J_{1}=1.90951(1)$ \cite{NSMa2018anomalous,Matsumoto2001}. (d) The phase diagram of the $J-Q_3$ model with $J=1$ and for which $Q_c=Q/J=1.49153$. }
\label{Fig2}
\end{figure}

{\it{\color{blue} Tower of states.-}}
Specifically, in the 2D square-lattice antiferromagnetic (AFM) Heisenberg model, the ground state spontaneously breaks the $SU(2)$ [or equivalently $SO(3)$] spin--rotation symmetry. At the thermodynamic limit, the ground states are infinitely degenerate, and the corresponding manifold is a 2D sphere. However, in a finite-size calculation, the symmetry is always restored, and instead, we have a unique ground state and a “tower” in the excited state~\cite{anderson1952approximate,lhuillier2005frustratedquantummagnets,wietek2017studying}. The tower can be described by an effective Hamiltonian~\cite{metlitski2015entanglement},
\begin{equation}
H_\mathrm{tower} = \frac{c^2 \mathbf{S}^2}{2\rho_S L^d},
\end{equation}
where $c$ is the spin wave velocity, $\mathbf{S}$ is the total spin of the system, $\rho_S$ is the spin stiffness, $L$ is the linear size of the system, and $d$ is the dimension of the lattice. For more general cases~\cite{Penc2003PRBSU4,Toth2010PRLSU3,Corboz2011PRLSU4}, in the 2D square-lattice $SU(N)$ Heisenberg model, the square of total spin $\mathbf{S}^2$ should be generalized to the quadratic Casimir operator $C_2 = \mathbf{J}^2$, where the components of $\mathbf{J}$ are the generators of the Lie algebra $\mathfrak{su}(N)$.
This generalization also holds for the $SO(N)$ group and $\mathfrak{so}(N)$ algebra, and we find that the low-lying levels correspond to the fully symmetric tensor representation labeled in the 1-row Young diagram, in which the $C_2$ eigenvalues can be evaluated by $\mathcal{J}(\mathcal{J}+N-2)$, where $\mathcal{J}$ is the number of boxes in the 1-row Young diagram and physically corresponds to the total $SO(N)$-spin of the considered level. In summary, we then arrive at the general form of the TOS~\cite{Hasenfratz1993} in $SO(N)$-SSB states,
\begin{equation}
E_\mathcal{J}(L)-E_0(L)=\frac{c^2 \mathbf{J}^2}{2\rho_\mathcal{J} L^d}=\frac{c^2 \mathcal{J}(\mathcal{J}+N-2)}{2\rho_\mathcal{J} L^d},
\label{eq3}
\end{equation}
where $\rho_\mathcal{J}$ is the $SO(N)$ spin stiffness. Note that the structure of the TOS contains the $N$ of the related group; thus, it probes the symmetry.

{\it{\color{blue} ES of (2+1)D $O(3)$ criticality.-}}
To demonstrate that the TOS of ES can reflect the symmetry of the critical point through reducing $T_E$, we use the spin-1/2 columnar dimerized Heisenberg model on square lattice with (2+1)D $O(3)$ criticality to visualize our concept. The Hamiltonian is given by
\begin{equation}
H=J_1\sum_{\langle ij \rangle}S_iS_j+J_2\sum_{\langle ij \rangle}S_iS_j
\label{HJ1J2}
\end{equation}
where $\langle ij \rangle$ denotes the nearest-neighbor bonds and  $J_1$ and $J_2$ are the coupling strengths of the thin and thick bonds, respectively, as shown in Fig.\ref{Fig2} (a). The corresponding ground-state phase diagram (Fig.\ref{Fig2} (c)) has been accurately determined in previous studies \cite{Matsumoto2001,ding2018engineering}.

As shown in Fig. \ref{fig1}, $J=J_2/J_1$ is the tunable parameter of the OH in Eq. (\ref{HJ1J2}). The red region denotes the N\'eel phase with $O(3)$ symmetry breaking, and the blue region denotes the dimer phase. The yellow point here is a (2+1)D $O(3)$ critical point. We expect that the ES at $J_c$ reflects the continuous $O(3)$ SSB at $T_E=0$. 
Here, we choose a 1D chain in the 2D lattice with a periodic boundary condition (PBC) as the subsystem $A$, which requires few computer memory for our quantum Monte Carlo (QMC) calculations, as shown in Fig.\ref{Fig2} (a). The additional reason is that the subregion should be cornerless which can hold the TOS \cite{Alba2013entanglement,Kolley2013entanglement,rademaker2015tower}.
{It's worth noting that the subsystem is not necessary being a chain, cornerless cut is enough, such as half-cut, because the general phase diagram of EH in Fig.\ref{fig1} also holds in this case. In the main text, we focus on the QMC simulations in which we choose the subsystem as a 1D chain. We also use density matrix renormalization group (DMRG) to demonstrate that this idea also works for tensor-like methods and half-cut, the results can be found in the Supplemental Material (SM)\cite{sm}.}

\begin{figure}[h!]
    \centering
    \begin{minipage}[c]{0.45\textwidth}
    \includegraphics[width=\textwidth]{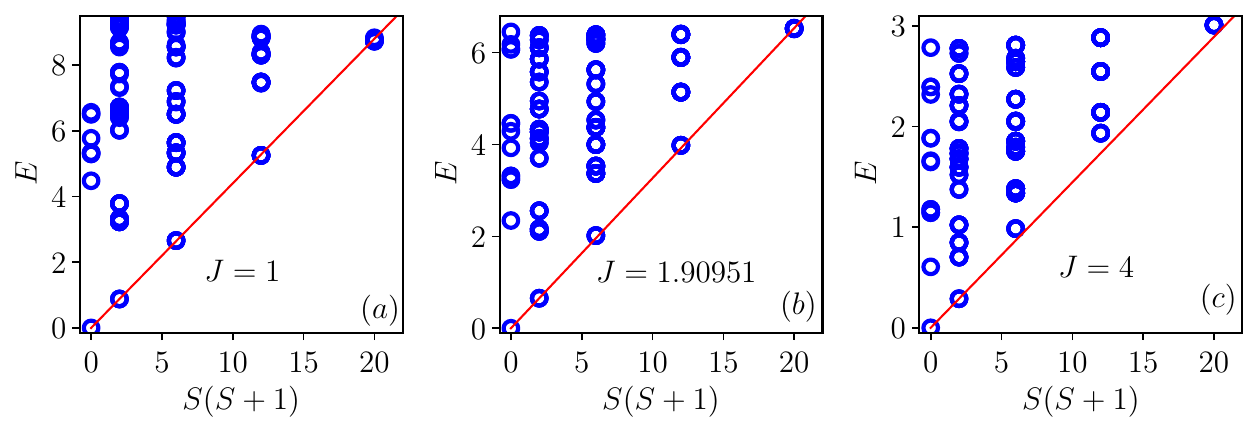}
    \end{minipage}
\caption{Entanglement spectrum of the columnar Heisenberg model with $L_x=L_y=8$. The subsystem is chosen as a chain $L^{sub}_x=8$ with a PBC and various $J=J_2/J_1$. The QCP is $J_c=J_2/J_1=1.90951$. In (a) and (b), the tower is proportional to $S(S+1)$, which reflects the $O(3)$ SSB here. All the error bars are smaller than the data symbols.}
	\label{fig:J1J2_spec}
\end{figure}

Through the newly proposed {reduced density matrix quantum Monte Carlo (RDM-QMC)} method~\cite{mao2025sampling}, the RDM $\rho_A$ is sampled and then diagonalized to obtain the ES with a limited computational load. The key idea of the method is sampling the path integral with an open boundary in imaginary time and using the frequency of the paths with same initial/final states to approach the corresponding element of RDM. {In the following, we choose $\beta=100$ which is a temperature low enough for the gapped region we studied. The sampling number is $10^{10}$ which is enough since the low energy levels converge first.} In the SM\cite{sm}, we give the process to determine $N$ by the root-mean-square error (RMSE).
Fig.\ref{fig:J1J2_spec} shows that the ES at both $J=J_c=1.90951$ and $J=1$ displays a structure of $O(3)$ SSB in which the tower is proportional to $S(S+1)$, where $S$ is the total spin. The ES here is completely different from the energy spectrum at a critical point, which is a CFT tower~\cite{francesco2012conformal,zhu2023uncovering}.
One may be concerned with whether a finite size effect makes the ES at $J_c$ seem similar to the one in the N\'eel phase nearby. Later, we will show that if the symmetry of the critical point is different from that for other points, the TOS reflects only the SSB of the critical point. In other words, the TOS at $J_c$ exactly reflects the symmetry of the phase transition point. In the SM\cite{sm}, we analyzed the robustness of the structure of TOS near the critical point and it has no obvious finite size effect. 

{\it{\color{blue} SO(3) spin $S$ reveals the TOS of the SO(N) superspin $\mathcal{J}$.-}}
For a more general $SO(N)$ TOS, in practice, we can access only some of the degrees of freedom and the corresponding quantum numbers of the subgroup. For example, the total spin $S$ of the $SO(3)$ subgroup is the real spin we can measure directly.
In these cases, the irreducible representations (IREPs) of the original $SO(N)$ group usually become reducible for the $SO(3)$ subgroup. We can then still identify the $SO(N)$ IREPs in the $SO(3)$ spin sectors [c.f. TABLE~\ref{tab:tab1} for the cases of $SO(5)$ IREPs decomposed into direct sums of $SO(3)$ IREPs]. 
For example, the $SO(5)$ $\mathcal{J}=1$ IREP splits as $\mathbf{5}_{\mathcal{J}=1} \rightarrow 2_{S=0} \otimes 3_{S=1}$, and the $\mathcal{J}=2$ IREP splits as $\mathbf{14}_{\mathcal{J}=2} \rightarrow 3_{S=0} \otimes 6_{S=1} \otimes 5_{S=2}$, 
where the notation $\mathbf{M}_{\mathcal{J}}$ refers to the degeneracy $\mathbf{M}$ in the $SO(5)$ spin sector $\mathcal{J}$ and $M_{S}$ refers to the degeneracy $M$ in the $SO(3)$ spin sector $S$.

Alternatively, we offer a {\it Poorman's approach} by simply replacing the $SO(N)$ spin $\mathcal{J}$ with $SO(3)$ spin $S$ in the formula Eq.~\ref{eq3}), i.e.,
\begin{equation}
E_S(L)-E_0(L)\propto S(S+N-2).
\label{eq4}
\end{equation}
This means that the proportional relation still holds for a subgroup (real spin). This can be justified by the fact that the maximum $S$ a state can reach is $\mathcal{J}$, and the maximum $S$ levels in the spectrum are located at the precise locations of the corresponding $SO(N)$ IREPs.

\begin{table}[h!]
\newcolumntype{M}[1]{>{\centering\arraybackslash}m{#1}}
\ytableausetup{boxsize=.55em}
\caption{The Young diagrams of different SO(5) irreducible representations (denoted as IREPs) 
and the corresponding state degeneracies in different sectors with total SO(3) spin $S$. 
}
\begin{tabular}{M{1.5cm}|M{1.5cm}|M{.7cm}M{.7cm}M{.7cm}M{.7cm}M{.7cm}}
    \toprule
    SO(5) & Young & \multicolumn{5}{c}{SO(3) spin sectors} \\
    %\cline{ 3 - 7 }
    IREP& diagram&0&1&2&3&4 
    \\[2pt]
    \hline
     {\bf1}$_{\mathcal{J}=0}$  &              &1& & & &  \\[6pt]
     {\bf5}$_{\mathcal{J}=1}$  &\ydiagram{1}  &2&3& & &  \\[6pt]
     {\bf14}$_{\mathcal{J}=2}$&\ydiagram{2}  &3&6&5& & \\[6pt]
     {\bf30}$_{\mathcal{J}=3}$ &\ydiagram{3}  &4&9&10&7& \\[6pt]
     {\bf55}$_{\mathcal{J}=4}$ &\ydiagram{4}  &5&12&15&14&9 \\[6pt]
     \hline\hline
\end{tabular}
\label{tab:tab1}
\end{table}

{\it{\color{blue} ES in the $J-Q_3$ model with emergent $SO(5)$ symmetry.-}}
Directly, we explore the ES of the 2D square lattice $J-Q_3$ model with emergent $SO(5)$ symmetry by measuring real spin with $SO(3)$ symmetry (see Fig.\ref{Fig2}(a)). As shown in Fig.\ref{Fig2}(b), the subsystem is a linear subregion labeled $A$, and the remaining part is the environment $B$. We still sample the RDM via QMC and diagonalize the matrix to obtain the ES, as we did in the dimerized Heisenberg model.
The $J-Q_3$ Hamiltonian for the square lattice can be written as~\cite{Sandvik2007}
\begin{equation}
	\begin{split}
		H& = -J \sum_{\langle ij\rangle}P_{ij}-Q \sum_{\langle ijklmn\rangle}P_{ij}P_{kl}P_{mn}
	\end{split}
\end{equation}
where $P_{ij}=\frac{1}{4}-\mathbf{S}_i\cdot \mathbf{S}_j$.

Notably, the $S$ for assessing $SO(5)$ symmetry breaking should be the related $S$ operator in an $SO(5)$ group. However, the extra symmetry, except the $SO(3)$ symmetry of real spins, emerges from the original $Z_4$ VBS order at the DQCP of the $J-Q_3$ model to reconstruct the $SO(5)$ symmetry~\cite{nahumEmergent2015,takahashi2024so}. Therefore, we cannot measure the superspin of $SO(5)$ directly; instead, we use the $S$ of the $SO(3)$
spin here, i.e., the real spin, as mentioned in the above section.

\begin{figure}[h!]
    \centering
    \begin{minipage}[c]{0.49\textwidth}
    \includegraphics[width=\textwidth]{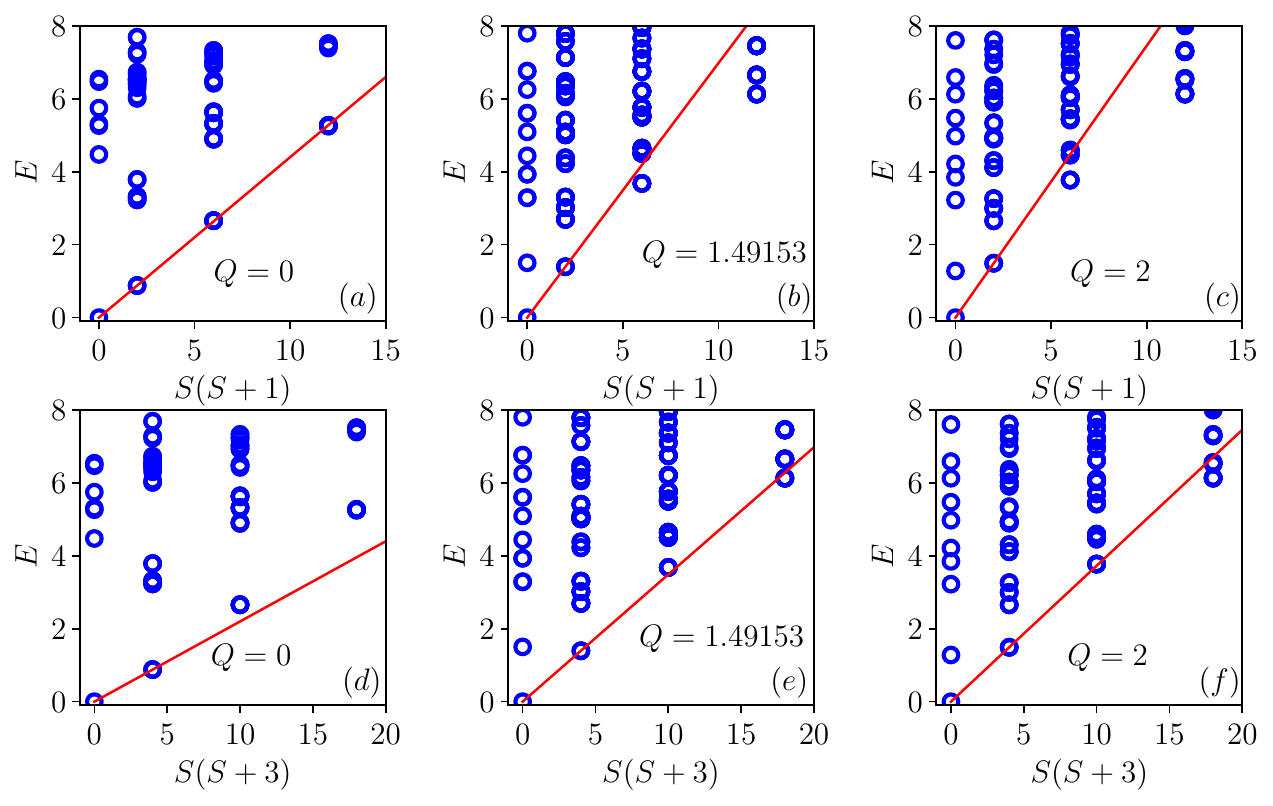}
    \end{minipage}
  
    \caption{Entanglement spectrum of the $J-Q_3$ square lattice model with $L_x=L_y=8$. The subsystem is chosen as a chain with size $L_x^{sub}=8$ and various $Q$ values. We choose $J=1$ as the energy unit, and the results are obtained {by RDM-QMC method}. The different linear relationships $S(S+N-2)$ reflect different degrees of symmetry breaking.  All the error bars are smaller than the data symbols.}
	\label{fig:JQ_spec}
\end{figure}

The critical point between the N\'eel and VBS phases is approximately $Q/J=1.49153$~\cite{takahashi2024so,wangScaling2022,zhaoScaling2022}. According to the knowledge of TOS, the ES is proportional to $S(S+N-2)$ when the $O(N)$ symmetry is broken in the ground state. As shown in Fig.~\ref{fig:JQ_spec}, the transverse axes of the subfigures in the upper row are labeled $S(S+1)$. When $Q=0$, the ground state is an AFM N\'eel order with $O(3)$ SSB, and the ES displays a perfect linear dispersion with $S(S+1)$. Moreover, we set the transverse axis in the lower row as $S(S+3)$, and the dispersion of ES would be linear if the ground state is $SO(5)$ SSB. The subfigures show that the ES around the DQCP becomes linear, which reveals the emergent $SO(5)$ symmetry at the DQCP.

We note that the $SO(5)$ DQCP has recently been considered a weakly first-order phase transition ~\cite{takahashi2024so,zhaoScaling2022,deng2024diagnosing,zhou2024mathrmso5,song2023deconfined,wangScaling2022}, but our method remains valid even if there is an approximate DQCP with emergent symmetry. The reason is straight that the cooling of $T_E$ only strengthen the SSB but will not change the symmetry.

{\it{\color{blue} Checkerboard $J-Q$ model with emergent $O(4)$ symmetry.-}}
Similarly, the Hamiltonian of the checkerboard $J$–$Q$ model (CBJQ) on a 2D square lattice can be defined by singlet projection operators $P_{ij}=1/4-S_{i}S_{j}$~\cite{BWZhao2019,wang2024probing}
\begin{equation}
H=-J\sum_{\langle ij \rangle}P_{ij}-Q\sum_{ijkl\in \Box^{\prime}}P_{ij}P_{kl}
\label{eq:cbjq}
\end{equation}
where all indicated site pairs comprise nearest neighbors and $\Box^{\prime}$ denotes same-colored squares on the black and white checkerboard.

The ground-state phase diagram is accurately determined via the quantum Monte Carlo method~\cite{BWZhao2019}. For $Q\to \infty$, the model is in the twofold degenerate plaquette-singlet solid phase, and for $Q\to 0$, the system is in the Néel phase. At the first-order phase transition point $Q/J\sim 4.598$~\cite{BWZhao2019}, the model is ordered by breaking emergent $O(4)$ symmetry~\cite{BWZhao2019}.

\begin{figure}[h!]
    \centering
    \begin{minipage}[c]{0.45\textwidth}
    \includegraphics[width=\textwidth]{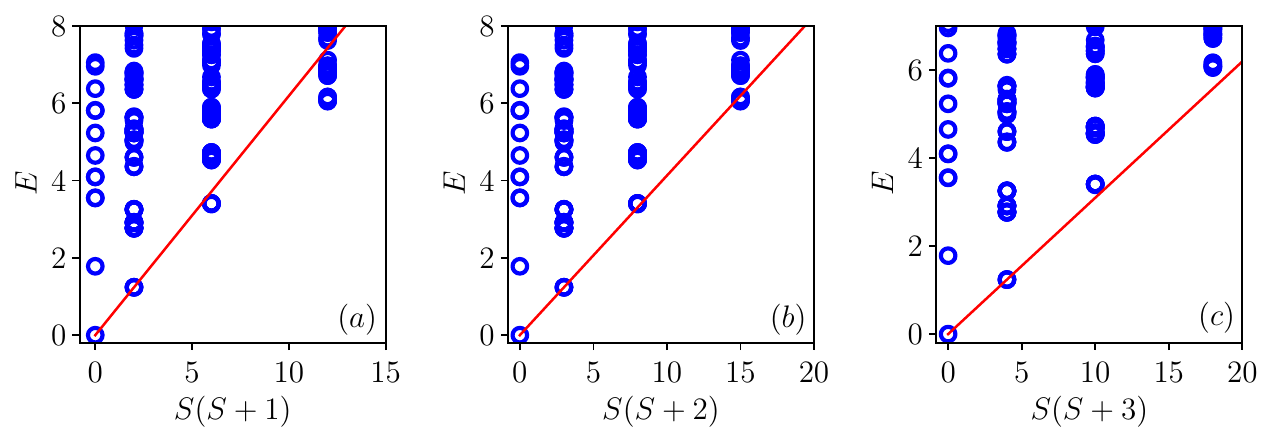}
    \end{minipage}
\caption{Entanglement spectrum of the checkerboard $J-Q$ model with $L_x=L_y=8$ at the first-order phase transition point. The subsystem is chosen as a chain of size $L_x^{sub}=8$. The linear relation $S(S+2)$ is associated with the $O(4)$ SSB here.  All the error bars are smaller than the data symbols.}
	\label{fig:cbjq_spec}
\end{figure}

At the first-order phase transition point with an emergent $O(4)$ SSB, we also calculate the ES and draw it versus $S(S+1)$, $S(S+2)$ and $S(S+3)$, where $S$ is the real spin with $SO(3)$ symmetry. Fig.\ref{fig:cbjq_spec} shows that the TOS of ES clearly supports the emergent $O(4)$ symmetry (breaking) in the CBJQ model.

{\it{\color{blue} Conclusion.-}}
We propose a general solution to numerically detect the underlying symmetry of a phase transition point without any prior knowledge of the corresponding low-energy effective field theory.
Thanks to the extra temperature axis introduced by EH, we calculate the ES of the system, the actual TOS of which reflects the SSB of related symmetry at the critical point due to the reduction of the extra temperature.
The good properties of the TOS are perfectly preserved at finite size, making it possible to assess SSB intuitively with limited computational resources even when the correlation length of the system is large. For emergent symmetry, our scheme does not require measuring the related superspin, and only the real spin is needed to demarcate symmetry on the basis of the projection principle in group theory.
The numerical scheme is easily implemented in both QMC and tensor networks without optimizing in a large parameter space~\cite{yang2023detecting}; several (2+1)D quantum phase transitions were explored as examples.
{Once the subsystem is cornerless, this scheme works well in either QMC or tensor-like methods (DMRG results are displayed in the SM\cite{sm}). For convinience, we choose subsystem as a 1D chain in QMC and half-cut in DMRG.}
This low-cost and low-technical-threshold method will greatly enhance studies of the quantum criticalities of unknown systems and aid in constructing low-energy effective field theories for these systems.

{\it{\color{blue} Acknowledgments.-}}
We would like to thank Zehui Deng, Shuai Yin, Rui-Zhen Huang, Wei Zhu, Chengxiang Ding, Fabien Alet and Sylvain Capponi for fruitful discussions. BBM acknowledges the Natural Science Foundation of Shandong Province, China (Grant No. ZR2024QA194). ZW thanks the China Postdoctoral Science Foundation under Grant No. 2024M752898. BBC is supported by the Fundamental Research Funds for the Central Universities.  The work is supported by the Scientific Research Project (No.WU2025B011) and the Start-up Funding of Westlake University.
The authors also acknowledge the HPC Centre of Westlake University and Beijing PARATERA Tech Co., Ltd., for providing HPC resources. 

{\it{\color{blue} Data availability.-}}
The data that support the findings of this article are openly available\cite{dataset}.

\bibliography{tos}

\begin{thebibliography}{83}%
\makeatletter
\providecommand \@ifxundefined [1]{%
 \@ifx{#1\undefined}
}%
\providecommand \@ifnum [1]{%
 \ifnum #1\expandafter \@firstoftwo
 \else \expandafter \@secondoftwo
 \fi
}%
\providecommand \@ifx [1]{%
 \ifx #1\expandafter \@firstoftwo
 \else \expandafter \@secondoftwo
 \fi
}%
\providecommand \natexlab [1]{#1}%
\providecommand \enquote  [1]{``#1''}%
\providecommand \bibnamefont  [1]{#1}%
\providecommand \bibfnamefont [1]{#1}%
\providecommand \citenamefont [1]{#1}%
\providecommand \href@noop [0]{\@secondoftwo}%
\providecommand \href [0]{\begingroup \@sanitize@url \@href}%
\providecommand \@href[1]{\@@startlink{#1}\@@href}%
\providecommand \@@href[1]{\endgroup#1\@@endlink}%
\providecommand \@sanitize@url [0]{\catcode `\\12\catcode `\$12\catcode `\&12\catcode `\#12\catcode `\^12\catcode `\_12\catcode `\%12\relax}%
\providecommand \@@startlink[1]{}%
\providecommand \@@endlink[0]{}%
\providecommand \url  [0]{\begingroup\@sanitize@url \@url }%
\providecommand \@url [1]{\endgroup\@href {#1}{\urlprefix }}%
\providecommand \urlprefix  [0]{URL }%
\providecommand \Eprint [0]{\href }%
\providecommand \doibase [0]{https://doi.org/}%
\providecommand \selectlanguage [0]{\@gobble}%
\providecommand \bibinfo  [0]{\@secondoftwo}%
\providecommand \bibfield  [0]{\@secondoftwo}%
\providecommand \translation [1]{[#1]}%
\providecommand \BibitemOpen [0]{}%
\providecommand \bibitemStop [0]{}%
\providecommand \bibitemNoStop [0]{.\EOS\space}%
\providecommand \EOS [0]{\spacefactor3000\relax}%
\providecommand \BibitemShut  [1]{\csname bibitem#1\endcsname}%
\let\auto@bib@innerbib\@empty
%</preamble>
\bibitem [{\citenamefont {Sachdev}(1999)}]{sachdev1999quantum}%
  \BibitemOpen
  \bibfield  {author} {\bibinfo {author} {\bibfnamefont {S.}~\bibnamefont {Sachdev}},\ }\bibfield  {title} {\bibinfo {title} {Quantum phase transitions},\ }\href@noop {} {\bibfield  {journal} {\bibinfo  {journal} {Physics world}\ }\textbf {\bibinfo {volume} {12}},\ \bibinfo {pages} {33} (\bibinfo {year} {1999})}\BibitemShut {NoStop}%
\bibitem [{\citenamefont {Wannier}(1987)}]{wannier1987statistical}%
  \BibitemOpen
  \bibfield  {author} {\bibinfo {author} {\bibfnamefont {G.~H.}\ \bibnamefont {Wannier}},\ }\href@noop {} {\emph {\bibinfo {title} {Statistical physics}}}\ (\bibinfo  {publisher} {Courier Corporation},\ \bibinfo {year} {1987})\BibitemShut {NoStop}%
\bibitem [{\citenamefont {Girvin}\ and\ \citenamefont {Yang}(2019)}]{girvin2019modern}%
  \BibitemOpen
  \bibfield  {author} {\bibinfo {author} {\bibfnamefont {S.~M.}\ \bibnamefont {Girvin}}\ and\ \bibinfo {author} {\bibfnamefont {K.}~\bibnamefont {Yang}},\ }\href@noop {} {\emph {\bibinfo {title} {Modern condensed matter physics}}}\ (\bibinfo  {publisher} {Cambridge University Press},\ \bibinfo {year} {2019})\BibitemShut {NoStop}%
\bibitem [{\citenamefont {Zhao}\ \emph {et~al.}(2019)\citenamefont {Zhao}, \citenamefont {Weinberg},\ and\ \citenamefont {Sandvik}}]{BWZhao2019}%
  \BibitemOpen
  \bibfield  {author} {\bibinfo {author} {\bibfnamefont {B.}~\bibnamefont {Zhao}}, \bibinfo {author} {\bibfnamefont {P.}~\bibnamefont {Weinberg}},\ and\ \bibinfo {author} {\bibfnamefont {A.~W.}\ \bibnamefont {Sandvik}},\ }\bibfield  {title} {\bibinfo {title} {Symmetry-enhanced discontinuous phase transition in a two-dimensional quantum magnet},\ }\href {https://doi.org/10.1038/s41567-019-0484-x} {\bibfield  {journal} {\bibinfo  {journal} {Nature Physics}\ }\textbf {\bibinfo {volume} {15}},\ \bibinfo {pages} {678 } (\bibinfo {year} {2019})}\BibitemShut {NoStop}%
\bibitem [{\citenamefont {Yan}\ \emph {et~al.}(2021)\citenamefont {Yan}, \citenamefont {Zhou}, \citenamefont {Sylju\aa{}sen}, \citenamefont {Zhang}, \citenamefont {Yuan}, \citenamefont {Lou},\ and\ \citenamefont {Chen}}]{yan2021widely}%
  \BibitemOpen
  \bibfield  {author} {\bibinfo {author} {\bibfnamefont {Z.}~\bibnamefont {Yan}}, \bibinfo {author} {\bibfnamefont {Z.}~\bibnamefont {Zhou}}, \bibinfo {author} {\bibfnamefont {O.~F.}\ \bibnamefont {Sylju\aa{}sen}}, \bibinfo {author} {\bibfnamefont {J.}~\bibnamefont {Zhang}}, \bibinfo {author} {\bibfnamefont {T.}~\bibnamefont {Yuan}}, \bibinfo {author} {\bibfnamefont {J.}~\bibnamefont {Lou}},\ and\ \bibinfo {author} {\bibfnamefont {Y.}~\bibnamefont {Chen}},\ }\bibfield  {title} {\bibinfo {title} {Widely existing mixed phase structure of the quantum dimer model on a square lattice},\ }\href {https://doi.org/10.1103/PhysRevB.103.094421} {\bibfield  {journal} {\bibinfo  {journal} {Phys. Rev. B}\ }\textbf {\bibinfo {volume} {103}},\ \bibinfo {pages} {094421} (\bibinfo {year} {2021})}\BibitemShut {NoStop}%
\bibitem [{\citenamefont {Sun}\ \emph {et~al.}(2021{\natexlab{a}})\citenamefont {Sun}, \citenamefont {Ma}, \citenamefont {Zhao}, \citenamefont {Sandvik},\ and\ \citenamefont {Meng}}]{Sun_2021}%
  \BibitemOpen
  \bibfield  {author} {\bibinfo {author} {\bibfnamefont {G.}~\bibnamefont {Sun}}, \bibinfo {author} {\bibfnamefont {N.}~\bibnamefont {Ma}}, \bibinfo {author} {\bibfnamefont {B.}~\bibnamefont {Zhao}}, \bibinfo {author} {\bibfnamefont {A.~W.}\ \bibnamefont {Sandvik}},\ and\ \bibinfo {author} {\bibfnamefont {Z.~Y.}\ \bibnamefont {Meng}},\ }\bibfield  {title} {\bibinfo {title} {Emergent o(4) symmetry at the phase transition from plaquette-singlet to antiferromagnetic order in quasi-two-dimensional quantum magnets*},\ }\href {https://doi.org/10.1088/1674-1056/abf3b8} {\bibfield  {journal} {\bibinfo  {journal} {Chinese Physics B}\ }\textbf {\bibinfo {volume} {30}},\ \bibinfo {pages} {067505} (\bibinfo {year} {2021}{\natexlab{a}})}\BibitemShut {NoStop}%
\bibitem [{\citenamefont {Ding}\ \emph {et~al.}(2016)\citenamefont {Ding}, \citenamefont {Bl\"ote},\ and\ \citenamefont {Deng}}]{ding2016emergent}%
  \BibitemOpen
  \bibfield  {author} {\bibinfo {author} {\bibfnamefont {C.}~\bibnamefont {Ding}}, \bibinfo {author} {\bibfnamefont {H.~W.~J.}\ \bibnamefont {Bl\"ote}},\ and\ \bibinfo {author} {\bibfnamefont {Y.}~\bibnamefont {Deng}},\ }\bibfield  {title} {\bibinfo {title} {Emergent o($n$) symmetry in a series of three-dimensional potts models},\ }\href {https://doi.org/10.1103/PhysRevB.94.104402} {\bibfield  {journal} {\bibinfo  {journal} {Phys. Rev. B}\ }\textbf {\bibinfo {volume} {94}},\ \bibinfo {pages} {104402} (\bibinfo {year} {2016})}\BibitemShut {NoStop}%
\bibitem [{\citenamefont {Zhou}\ \emph {et~al.}(2025)\citenamefont {Zhou}, \citenamefont {Yan}, \citenamefont {Liu}, \citenamefont {Chen},\ and\ \citenamefont {Zhang}}]{Zhou_2025}%
  \BibitemOpen
  \bibfield  {author} {\bibinfo {author} {\bibfnamefont {Z.}~\bibnamefont {Zhou}}, \bibinfo {author} {\bibfnamefont {Z.}~\bibnamefont {Yan}}, \bibinfo {author} {\bibfnamefont {C.}~\bibnamefont {Liu}}, \bibinfo {author} {\bibfnamefont {Y.}~\bibnamefont {Chen}},\ and\ \bibinfo {author} {\bibfnamefont {X.-F.}\ \bibnamefont {Zhang}},\ }\bibfield  {title} {\bibinfo {title} {Quantum simulation of two-dimensional u(1) gauge theory in rydberg and rydberg-dressed atom arrays},\ }\href {https://doi.org/10.1088/0256-307X/42/5/053705} {\bibfield  {journal} {\bibinfo  {journal} {Chinese Physics Letters}\ }\textbf {\bibinfo {volume} {42}},\ \bibinfo {pages} {053705} (\bibinfo {year} {2025})}\BibitemShut {NoStop}%
\bibitem [{\citenamefont {Senthil}\ \emph {et~al.}(2004{\natexlab{a}})\citenamefont {Senthil}, \citenamefont {Vishwanath}, \citenamefont {Balents}, \citenamefont {Sachdev},\ and\ \citenamefont {Fisher}}]{senthilDeconfined2004}%
  \BibitemOpen
  \bibfield  {author} {\bibinfo {author} {\bibfnamefont {T.}~\bibnamefont {Senthil}}, \bibinfo {author} {\bibfnamefont {A.}~\bibnamefont {Vishwanath}}, \bibinfo {author} {\bibfnamefont {L.}~\bibnamefont {Balents}}, \bibinfo {author} {\bibfnamefont {S.}~\bibnamefont {Sachdev}},\ and\ \bibinfo {author} {\bibfnamefont {M.~P.~A.}\ \bibnamefont {Fisher}},\ }\bibfield  {title} {\bibinfo {title} {Deconfined quantum critical points},\ }\href {https://doi.org/10.1126/science.1091806} {\bibfield  {journal} {\bibinfo  {journal} {Science}\ }\textbf {\bibinfo {volume} {303}},\ \bibinfo {pages} {1490} (\bibinfo {year} {2004}{\natexlab{a}})}\BibitemShut {NoStop}%
\bibitem [{\citenamefont {Senthil}\ \emph {et~al.}(2004{\natexlab{b}})\citenamefont {Senthil}, \citenamefont {Balents}, \citenamefont {Sachdev}, \citenamefont {Vishwanath},\ and\ \citenamefont {Fisher}}]{senthilQuantum2004}%
  \BibitemOpen
  \bibfield  {author} {\bibinfo {author} {\bibfnamefont {T.}~\bibnamefont {Senthil}}, \bibinfo {author} {\bibfnamefont {L.}~\bibnamefont {Balents}}, \bibinfo {author} {\bibfnamefont {S.}~\bibnamefont {Sachdev}}, \bibinfo {author} {\bibfnamefont {A.}~\bibnamefont {Vishwanath}},\ and\ \bibinfo {author} {\bibfnamefont {M.~P.~A.}\ \bibnamefont {Fisher}},\ }\bibfield  {title} {\bibinfo {title} {Quantum criticality beyond the landau-ginzburg-wilson paradigm},\ }\href {https://doi.org/10.1103/PhysRevB.70.144407} {\bibfield  {journal} {\bibinfo  {journal} {Phys. Rev. B}\ }\textbf {\bibinfo {volume} {70}},\ \bibinfo {pages} {144407} (\bibinfo {year} {2004}{\natexlab{b}})}\BibitemShut {NoStop}%
\bibitem [{\citenamefont {Sandvik}(2007{\natexlab{a}})}]{sandvikEvidence2007}%
  \BibitemOpen
  \bibfield  {author} {\bibinfo {author} {\bibfnamefont {A.~W.}\ \bibnamefont {Sandvik}},\ }\bibfield  {title} {\bibinfo {title} {Evidence for deconfined quantum criticality in a two-dimensional heisenberg model with four-spin interactions},\ }\href {https://doi.org/10.1103/PhysRevLett.98.227202} {\bibfield  {journal} {\bibinfo  {journal} {Phys. Rev. Lett.}\ }\textbf {\bibinfo {volume} {98}},\ \bibinfo {pages} {227202} (\bibinfo {year} {2007}{\natexlab{a}})}\BibitemShut {NoStop}%
\bibitem [{\citenamefont {Nahum}\ \emph {et~al.}(2015{\natexlab{a}})\citenamefont {Nahum}, \citenamefont {Chalker}, \citenamefont {Serna}, \citenamefont {Ortu\~no},\ and\ \citenamefont {Somoza}}]{nahumDeconfined2015}%
  \BibitemOpen
  \bibfield  {author} {\bibinfo {author} {\bibfnamefont {A.}~\bibnamefont {Nahum}}, \bibinfo {author} {\bibfnamefont {J.~T.}\ \bibnamefont {Chalker}}, \bibinfo {author} {\bibfnamefont {P.}~\bibnamefont {Serna}}, \bibinfo {author} {\bibfnamefont {M.}~\bibnamefont {Ortu\~no}},\ and\ \bibinfo {author} {\bibfnamefont {A.~M.}\ \bibnamefont {Somoza}},\ }\bibfield  {title} {\bibinfo {title} {Deconfined quantum criticality, scaling violations, and classical loop models},\ }\href {https://doi.org/10.1103/PhysRevX.5.041048} {\bibfield  {journal} {\bibinfo  {journal} {Phys. Rev. X}\ }\textbf {\bibinfo {volume} {5}},\ \bibinfo {pages} {041048} (\bibinfo {year} {2015}{\natexlab{a}})}\BibitemShut {NoStop}%
\bibitem [{\citenamefont {Qin}\ \emph {et~al.}(2017)\citenamefont {Qin}, \citenamefont {He}, \citenamefont {You}, \citenamefont {Lu}, \citenamefont {Sen}, \citenamefont {Sandvik}, \citenamefont {Xu},\ and\ \citenamefont {Meng}}]{qinDuality2017}%
  \BibitemOpen
  \bibfield  {author} {\bibinfo {author} {\bibfnamefont {Y.~Q.}\ \bibnamefont {Qin}}, \bibinfo {author} {\bibfnamefont {Y.-Y.}\ \bibnamefont {He}}, \bibinfo {author} {\bibfnamefont {Y.-Z.}\ \bibnamefont {You}}, \bibinfo {author} {\bibfnamefont {Z.-Y.}\ \bibnamefont {Lu}}, \bibinfo {author} {\bibfnamefont {A.}~\bibnamefont {Sen}}, \bibinfo {author} {\bibfnamefont {A.~W.}\ \bibnamefont {Sandvik}}, \bibinfo {author} {\bibfnamefont {C.}~\bibnamefont {Xu}},\ and\ \bibinfo {author} {\bibfnamefont {Z.~Y.}\ \bibnamefont {Meng}},\ }\bibfield  {title} {\bibinfo {title} {Duality between the deconfined quantum-critical point and the bosonic topological transition},\ }\href {https://doi.org/10.1103/PhysRevX.7.031052} {\bibfield  {journal} {\bibinfo  {journal} {Phys. Rev. X}\ }\textbf {\bibinfo {volume} {7}},\ \bibinfo {pages} {031052} (\bibinfo {year} {2017})}\BibitemShut {NoStop}%
\bibitem [{\citenamefont {Wang}\ \emph {et~al.}(2017{\natexlab{a}})\citenamefont {Wang}, \citenamefont {Nahum}, \citenamefont {Metlitski}, \citenamefont {Xu},\ and\ \citenamefont {Senthil}}]{wangDeconfined2017}%
  \BibitemOpen
  \bibfield  {author} {\bibinfo {author} {\bibfnamefont {C.}~\bibnamefont {Wang}}, \bibinfo {author} {\bibfnamefont {A.}~\bibnamefont {Nahum}}, \bibinfo {author} {\bibfnamefont {M.~A.}\ \bibnamefont {Metlitski}}, \bibinfo {author} {\bibfnamefont {C.}~\bibnamefont {Xu}},\ and\ \bibinfo {author} {\bibfnamefont {T.}~\bibnamefont {Senthil}},\ }\bibfield  {title} {\bibinfo {title} {Deconfined quantum critical points: Symmetries and dualities},\ }\href {https://doi.org/10.1103/PhysRevX.7.031051} {\bibfield  {journal} {\bibinfo  {journal} {Phys. Rev. X}\ }\textbf {\bibinfo {volume} {7}},\ \bibinfo {pages} {031051} (\bibinfo {year} {2017}{\natexlab{a}})}\BibitemShut {NoStop}%
\bibitem [{\citenamefont {Senthil}(2023)}]{senthilDeconfined2023}%
  \BibitemOpen
  \bibfield  {author} {\bibinfo {author} {\bibfnamefont {T.}~\bibnamefont {Senthil}},\ }\href@noop {} {\bibinfo {title} {Deconfined quantum critical points: a review}} (\bibinfo {year} {2023}),\ \Eprint {https://arxiv.org/abs/2306.12638} {arXiv:2306.12638 [cond-mat.str-el]} \BibitemShut {NoStop}%
\bibitem [{\citenamefont {Nahum}\ \emph {et~al.}(2015{\natexlab{b}})\citenamefont {Nahum}, \citenamefont {Serna}, \citenamefont {Chalker}, \citenamefont {Ortu\~no},\ and\ \citenamefont {Somoza}}]{nahumEmergent2015}%
  \BibitemOpen
  \bibfield  {author} {\bibinfo {author} {\bibfnamefont {A.}~\bibnamefont {Nahum}}, \bibinfo {author} {\bibfnamefont {P.}~\bibnamefont {Serna}}, \bibinfo {author} {\bibfnamefont {J.~T.}\ \bibnamefont {Chalker}}, \bibinfo {author} {\bibfnamefont {M.}~\bibnamefont {Ortu\~no}},\ and\ \bibinfo {author} {\bibfnamefont {A.~M.}\ \bibnamefont {Somoza}},\ }\bibfield  {title} {\bibinfo {title} {Emergent so(5) symmetry at the n\'eel to valence-bond-solid transition},\ }\href {https://doi.org/10.1103/PhysRevLett.115.267203} {\bibfield  {journal} {\bibinfo  {journal} {Phys. Rev. Lett.}\ }\textbf {\bibinfo {volume} {115}},\ \bibinfo {pages} {267203} (\bibinfo {year} {2015}{\natexlab{b}})}\BibitemShut {NoStop}%
\bibitem [{\citenamefont {Tanaka}\ and\ \citenamefont {Hu}(2005)}]{tanaka2005many}%
  \BibitemOpen
  \bibfield  {author} {\bibinfo {author} {\bibfnamefont {A.}~\bibnamefont {Tanaka}}\ and\ \bibinfo {author} {\bibfnamefont {X.}~\bibnamefont {Hu}},\ }\bibfield  {title} {\bibinfo {title} {Many-body spin berry phases emerging from the $\ensuremath{\pi}$-flux state: Competition between antiferromagnetism and the valence-bond-solid state},\ }\href {https://doi.org/10.1103/PhysRevLett.95.036402} {\bibfield  {journal} {\bibinfo  {journal} {Phys. Rev. Lett.}\ }\textbf {\bibinfo {volume} {95}},\ \bibinfo {pages} {036402} (\bibinfo {year} {2005})}\BibitemShut {NoStop}%
\bibitem [{\citenamefont {Senthil}\ and\ \citenamefont {Fisher}(2006{\natexlab{a}})}]{senthil2006competing}%
  \BibitemOpen
  \bibfield  {author} {\bibinfo {author} {\bibfnamefont {T.}~\bibnamefont {Senthil}}\ and\ \bibinfo {author} {\bibfnamefont {M.~P.~A.}\ \bibnamefont {Fisher}},\ }\bibfield  {title} {\bibinfo {title} {Competing orders, nonlinear sigma models, and topological terms in quantum magnets},\ }\href {https://doi.org/10.1103/PhysRevB.74.064405} {\bibfield  {journal} {\bibinfo  {journal} {Phys. Rev. B}\ }\textbf {\bibinfo {volume} {74}},\ \bibinfo {pages} {064405} (\bibinfo {year} {2006}{\natexlab{a}})}\BibitemShut {NoStop}%
\bibitem [{\citenamefont {Liu}\ \emph {et~al.}(2024{\natexlab{a}})\citenamefont {Liu}, \citenamefont {Gong}, \citenamefont {Chen},\ and\ \citenamefont {Gu}}]{liu2024emergent}%
  \BibitemOpen
  \bibfield  {author} {\bibinfo {author} {\bibfnamefont {W.-Y.}\ \bibnamefont {Liu}}, \bibinfo {author} {\bibfnamefont {S.-S.}\ \bibnamefont {Gong}}, \bibinfo {author} {\bibfnamefont {W.-Q.}\ \bibnamefont {Chen}},\ and\ \bibinfo {author} {\bibfnamefont {Z.-C.}\ \bibnamefont {Gu}},\ }\bibfield  {title} {\bibinfo {title} {Emergent symmetry in quantum phase transition: From deconfined quantum critical point to gapless quantum spin liquid},\ }\href {https://doi.org/https://doi.org/10.1016/j.scib.2023.11.057} {\bibfield  {journal} {\bibinfo  {journal} {Science Bulletin}\ }\textbf {\bibinfo {volume} {69}},\ \bibinfo {pages} {190} (\bibinfo {year} {2024}{\natexlab{a}})}\BibitemShut {NoStop}%
\bibitem [{\citenamefont {Huang}\ \emph {et~al.}(2019)\citenamefont {Huang}, \citenamefont {Lu}, \citenamefont {You}, \citenamefont {Meng},\ and\ \citenamefont {Xiang}}]{huang2019emergent}%
  \BibitemOpen
  \bibfield  {author} {\bibinfo {author} {\bibfnamefont {R.-Z.}\ \bibnamefont {Huang}}, \bibinfo {author} {\bibfnamefont {D.-C.}\ \bibnamefont {Lu}}, \bibinfo {author} {\bibfnamefont {Y.-Z.}\ \bibnamefont {You}}, \bibinfo {author} {\bibfnamefont {Z.~Y.}\ \bibnamefont {Meng}},\ and\ \bibinfo {author} {\bibfnamefont {T.}~\bibnamefont {Xiang}},\ }\bibfield  {title} {\bibinfo {title} {Emergent symmetry and conserved current at a one-dimensional incarnation of deconfined quantum critical point},\ }\href {https://doi.org/10.1103/PhysRevB.100.125137} {\bibfield  {journal} {\bibinfo  {journal} {Phys. Rev. B}\ }\textbf {\bibinfo {volume} {100}},\ \bibinfo {pages} {125137} (\bibinfo {year} {2019})}\BibitemShut {NoStop}%
\bibitem [{\citenamefont {Wang}\ \emph {et~al.}(2017{\natexlab{b}})\citenamefont {Wang}, \citenamefont {Nahum}, \citenamefont {Metlitski}, \citenamefont {Xu},\ and\ \citenamefont {Senthil}}]{wang2017deconfined}%
  \BibitemOpen
  \bibfield  {author} {\bibinfo {author} {\bibfnamefont {C.}~\bibnamefont {Wang}}, \bibinfo {author} {\bibfnamefont {A.}~\bibnamefont {Nahum}}, \bibinfo {author} {\bibfnamefont {M.~A.}\ \bibnamefont {Metlitski}}, \bibinfo {author} {\bibfnamefont {C.}~\bibnamefont {Xu}},\ and\ \bibinfo {author} {\bibfnamefont {T.}~\bibnamefont {Senthil}},\ }\bibfield  {title} {\bibinfo {title} {Deconfined quantum critical points: Symmetries and dualities},\ }\href {https://doi.org/10.1103/PhysRevX.7.031051} {\bibfield  {journal} {\bibinfo  {journal} {Phys. Rev. X}\ }\textbf {\bibinfo {volume} {7}},\ \bibinfo {pages} {031051} (\bibinfo {year} {2017}{\natexlab{b}})}\BibitemShut {NoStop}%
\bibitem [{\citenamefont {Ma}\ \emph {et~al.}(2019{\natexlab{a}})\citenamefont {Ma}, \citenamefont {You},\ and\ \citenamefont {Meng}}]{ma2019role}%
  \BibitemOpen
  \bibfield  {author} {\bibinfo {author} {\bibfnamefont {N.}~\bibnamefont {Ma}}, \bibinfo {author} {\bibfnamefont {Y.-Z.}\ \bibnamefont {You}},\ and\ \bibinfo {author} {\bibfnamefont {Z.~Y.}\ \bibnamefont {Meng}},\ }\bibfield  {title} {\bibinfo {title} {Role of noether's theorem at the deconfined quantum critical point},\ }\href {https://doi.org/10.1103/PhysRevLett.122.175701} {\bibfield  {journal} {\bibinfo  {journal} {Phys. Rev. Lett.}\ }\textbf {\bibinfo {volume} {122}},\ \bibinfo {pages} {175701} (\bibinfo {year} {2019}{\natexlab{a}})}\BibitemShut {NoStop}%
\bibitem [{\citenamefont {Senthil}(2024)}]{senthil2024deconfined}%
  \BibitemOpen
  \bibfield  {author} {\bibinfo {author} {\bibfnamefont {T.}~\bibnamefont {Senthil}},\ }\bibfield  {title} {\bibinfo {title} {Deconfined quantum critical points: a review},\ }\href@noop {} {\bibfield  {journal} {\bibinfo  {journal} {50 Years of the Renormalization Group: Dedicated to the Memory of Michael E Fisher}\ ,\ \bibinfo {pages} {169}} (\bibinfo {year} {2024})}\BibitemShut {NoStop}%
\bibitem [{\citenamefont {D'Emidio}\ and\ \citenamefont {Sandvik}(2024)}]{d2024entanglement}%
  \BibitemOpen
  \bibfield  {author} {\bibinfo {author} {\bibfnamefont {J.}~\bibnamefont {D'Emidio}}\ and\ \bibinfo {author} {\bibfnamefont {A.~W.}\ \bibnamefont {Sandvik}},\ }\bibfield  {title} {\bibinfo {title} {Entanglement entropy and deconfined criticality: Emergent so(5) symmetry and proper lattice bipartition},\ }\href {https://doi.org/10.1103/PhysRevLett.133.166702} {\bibfield  {journal} {\bibinfo  {journal} {Phys. Rev. Lett.}\ }\textbf {\bibinfo {volume} {133}},\ \bibinfo {pages} {166702} (\bibinfo {year} {2024})}\BibitemShut {NoStop}%
\bibitem [{\citenamefont {Vishwanath}\ and\ \citenamefont {Senthil}(2013)}]{ashvinsenthil}%
  \BibitemOpen
  \bibfield  {author} {\bibinfo {author} {\bibfnamefont {A.}~\bibnamefont {Vishwanath}}\ and\ \bibinfo {author} {\bibfnamefont {T.}~\bibnamefont {Senthil}},\ }\bibfield  {title} {\bibinfo {title} {Physics of three-dimensional bosonic topological insulators: Surface-deconfined criticality and quantized magnetoelectric effect},\ }\href {https://doi.org/10.1103/PhysRevX.3.011016} {\bibfield  {journal} {\bibinfo  {journal} {Phys. Rev. X}\ }\textbf {\bibinfo {volume} {3}},\ \bibinfo {pages} {011016} (\bibinfo {year} {2013})}\BibitemShut {NoStop}%
\bibitem [{\citenamefont {Senthil}\ and\ \citenamefont {Fisher}(2006{\natexlab{b}})}]{senthilfisher}%
  \BibitemOpen
  \bibfield  {author} {\bibinfo {author} {\bibfnamefont {T.}~\bibnamefont {Senthil}}\ and\ \bibinfo {author} {\bibfnamefont {M.~P.~A.}\ \bibnamefont {Fisher}},\ }\bibfield  {title} {\bibinfo {title} {Competing orders, nonlinear sigma models, and topological terms in quantum magnets},\ }\href {https://doi.org/10.1103/PhysRevB.74.064405} {\bibfield  {journal} {\bibinfo  {journal} {Phys. Rev. B}\ }\textbf {\bibinfo {volume} {74}},\ \bibinfo {pages} {064405} (\bibinfo {year} {2006}{\natexlab{b}})}\BibitemShut {NoStop}%
\bibitem [{\citenamefont {Ma}\ \emph {et~al.}(2019{\natexlab{b}})\citenamefont {Ma}, \citenamefont {You},\ and\ \citenamefont {Meng}}]{maRole2019}%
  \BibitemOpen
  \bibfield  {author} {\bibinfo {author} {\bibfnamefont {N.}~\bibnamefont {Ma}}, \bibinfo {author} {\bibfnamefont {Y.-Z.}\ \bibnamefont {You}},\ and\ \bibinfo {author} {\bibfnamefont {Z.~Y.}\ \bibnamefont {Meng}},\ }\bibfield  {title} {\bibinfo {title} {Role of noether's theorem at the deconfined quantum critical point},\ }\href {https://doi.org/10.1103/PhysRevLett.122.175701} {\bibfield  {journal} {\bibinfo  {journal} {Phys. Rev. Lett.}\ }\textbf {\bibinfo {volume} {122}},\ \bibinfo {pages} {175701} (\bibinfo {year} {2019}{\natexlab{b}})}\BibitemShut {NoStop}%
\bibitem [{\citenamefont {Sreejith}\ \emph {et~al.}(2019)\citenamefont {Sreejith}, \citenamefont {Powell},\ and\ \citenamefont {Nahum}}]{sreejithEmergent2019}%
  \BibitemOpen
  \bibfield  {author} {\bibinfo {author} {\bibfnamefont {G.~J.}\ \bibnamefont {Sreejith}}, \bibinfo {author} {\bibfnamefont {S.}~\bibnamefont {Powell}},\ and\ \bibinfo {author} {\bibfnamefont {A.}~\bibnamefont {Nahum}},\ }\bibfield  {title} {\bibinfo {title} {Emergent so(5) symmetry at the columnar ordering transition in the classical cubic dimer model},\ }\href {https://doi.org/10.1103/PhysRevLett.122.080601} {\bibfield  {journal} {\bibinfo  {journal} {Phys. Rev. Lett.}\ }\textbf {\bibinfo {volume} {122}},\ \bibinfo {pages} {080601} (\bibinfo {year} {2019})}\BibitemShut {NoStop}%
\bibitem [{\citenamefont {Ma}\ \emph {et~al.}(2018{\natexlab{a}})\citenamefont {Ma}, \citenamefont {Sun}, \citenamefont {You}, \citenamefont {Xu}, \citenamefont {Vishwanath}, \citenamefont {Sandvik},\ and\ \citenamefont {Meng}}]{maDynamics2018}%
  \BibitemOpen
  \bibfield  {author} {\bibinfo {author} {\bibfnamefont {N.}~\bibnamefont {Ma}}, \bibinfo {author} {\bibfnamefont {G.-Y.}\ \bibnamefont {Sun}}, \bibinfo {author} {\bibfnamefont {Y.-Z.}\ \bibnamefont {You}}, \bibinfo {author} {\bibfnamefont {C.}~\bibnamefont {Xu}}, \bibinfo {author} {\bibfnamefont {A.}~\bibnamefont {Vishwanath}}, \bibinfo {author} {\bibfnamefont {A.~W.}\ \bibnamefont {Sandvik}},\ and\ \bibinfo {author} {\bibfnamefont {Z.~Y.}\ \bibnamefont {Meng}},\ }\bibfield  {title} {\bibinfo {title} {Dynamical signature of fractionalization at a deconfined quantum critical point},\ }\href {https://doi.org/10.1103/PhysRevB.98.174421} {\bibfield  {journal} {\bibinfo  {journal} {Phys. Rev. B}\ }\textbf {\bibinfo {volume} {98}},\ \bibinfo {pages} {174421} (\bibinfo {year} {2018}{\natexlab{a}})}\BibitemShut {NoStop}%
\bibitem [{\citenamefont {Harada}\ \emph {et~al.}(2013)\citenamefont {Harada}, \citenamefont {Suzuki}, \citenamefont {Okubo}, \citenamefont {Matsuo}, \citenamefont {Lou}, \citenamefont {Watanabe}, \citenamefont {Todo},\ and\ \citenamefont {Kawashima}}]{haradaPossibility2013}%
  \BibitemOpen
  \bibfield  {author} {\bibinfo {author} {\bibfnamefont {K.}~\bibnamefont {Harada}}, \bibinfo {author} {\bibfnamefont {T.}~\bibnamefont {Suzuki}}, \bibinfo {author} {\bibfnamefont {T.}~\bibnamefont {Okubo}}, \bibinfo {author} {\bibfnamefont {H.}~\bibnamefont {Matsuo}}, \bibinfo {author} {\bibfnamefont {J.}~\bibnamefont {Lou}}, \bibinfo {author} {\bibfnamefont {H.}~\bibnamefont {Watanabe}}, \bibinfo {author} {\bibfnamefont {S.}~\bibnamefont {Todo}},\ and\ \bibinfo {author} {\bibfnamefont {N.}~\bibnamefont {Kawashima}},\ }\bibfield  {title} {\bibinfo {title} {Possibility of deconfined criticality in su($n$) heisenberg models at small $n$},\ }\href {https://doi.org/10.1103/PhysRevB.88.220408} {\bibfield  {journal} {\bibinfo  {journal} {Phys. Rev. B}\ }\textbf {\bibinfo {volume} {88}},\ \bibinfo {pages} {220408} (\bibinfo {year} {2013})}\BibitemShut {NoStop}%
\bibitem [{\citenamefont {Lou}\ \emph {et~al.}(2009)\citenamefont {Lou}, \citenamefont {Sandvik},\ and\ \citenamefont {Kawashima}}]{louvbsneel2009}%
  \BibitemOpen
  \bibfield  {author} {\bibinfo {author} {\bibfnamefont {J.}~\bibnamefont {Lou}}, \bibinfo {author} {\bibfnamefont {A.~W.}\ \bibnamefont {Sandvik}},\ and\ \bibinfo {author} {\bibfnamefont {N.}~\bibnamefont {Kawashima}},\ }\bibfield  {title} {\bibinfo {title} {Antiferromagnetic to valence-bond-solid transitions in two-dimensional $\text{SU}(n)$ heisenberg models with multispin interactions},\ }\href {https://doi.org/10.1103/PhysRevB.80.180414} {\bibfield  {journal} {\bibinfo  {journal} {Phys. Rev. B}\ }\textbf {\bibinfo {volume} {80}},\ \bibinfo {pages} {180414} (\bibinfo {year} {2009})}\BibitemShut {NoStop}%
\bibitem [{\citenamefont {Liu}\ \emph {et~al.}(2019)\citenamefont {Liu}, \citenamefont {Wang}, \citenamefont {Sato}, \citenamefont {Hohenadler}, \citenamefont {Wang}, \citenamefont {Guo},\ and\ \citenamefont {Assaad}}]{liuSuperconductivity2019}%
  \BibitemOpen
  \bibfield  {author} {\bibinfo {author} {\bibfnamefont {Y.}~\bibnamefont {Liu}}, \bibinfo {author} {\bibfnamefont {Z.}~\bibnamefont {Wang}}, \bibinfo {author} {\bibfnamefont {T.}~\bibnamefont {Sato}}, \bibinfo {author} {\bibfnamefont {M.}~\bibnamefont {Hohenadler}}, \bibinfo {author} {\bibfnamefont {C.}~\bibnamefont {Wang}}, \bibinfo {author} {\bibfnamefont {W.}~\bibnamefont {Guo}},\ and\ \bibinfo {author} {\bibfnamefont {F.~F.}\ \bibnamefont {Assaad}},\ }\bibfield  {title} {\bibinfo {title} {Superconductivity from the condensation of topological defects in a quantum spin-hall insulator},\ }\href {https://doi.org/10.1038/s41467-019-10372-0} {\bibfield  {journal} {\bibinfo  {journal} {Nat. Commun.}\ }\textbf {\bibinfo {volume} {10}},\ \bibinfo {pages} {1} (\bibinfo {year} {2019})}\BibitemShut {NoStop}%
\bibitem [{\citenamefont {Da~Liao}\ \emph {et~al.}(2022)\citenamefont {Da~Liao}, \citenamefont {Xu}, \citenamefont {Meng},\ and\ \citenamefont {Qi}}]{liaoDiracI2022}%
  \BibitemOpen
  \bibfield  {author} {\bibinfo {author} {\bibfnamefont {Y.}~\bibnamefont {Da~Liao}}, \bibinfo {author} {\bibfnamefont {X.~Y.}\ \bibnamefont {Xu}}, \bibinfo {author} {\bibfnamefont {Z.~Y.}\ \bibnamefont {Meng}},\ and\ \bibinfo {author} {\bibfnamefont {Y.}~\bibnamefont {Qi}},\ }\bibfield  {title} {\bibinfo {title} {Dirac fermions with plaquette interactions. i. su(2) phase diagram with gross-neveu and deconfined quantum criticalities},\ }\href {https://doi.org/10.1103/PhysRevB.106.075111} {\bibfield  {journal} {\bibinfo  {journal} {Phys. Rev. B}\ }\textbf {\bibinfo {volume} {106}},\ \bibinfo {pages} {075111} (\bibinfo {year} {2022})}\BibitemShut {NoStop}%
\bibitem [{\citenamefont {Shao}\ \emph {et~al.}(2016)\citenamefont {Shao}, \citenamefont {Guo},\ and\ \citenamefont {Sandvik}}]{shaoQuantum2016}%
  \BibitemOpen
  \bibfield  {author} {\bibinfo {author} {\bibfnamefont {H.}~\bibnamefont {Shao}}, \bibinfo {author} {\bibfnamefont {W.}~\bibnamefont {Guo}},\ and\ \bibinfo {author} {\bibfnamefont {A.~W.}\ \bibnamefont {Sandvik}},\ }\bibfield  {title} {\bibinfo {title} {Quantum criticality with two length scales},\ }\href {https://doi.org/10.1126/science.aad5007} {\bibfield  {journal} {\bibinfo  {journal} {Science}\ }\textbf {\bibinfo {volume} {352}},\ \bibinfo {pages} {213} (\bibinfo {year} {2016})}\BibitemShut {NoStop}%
\bibitem [{\citenamefont {Zhang}\ \emph {et~al.}(2018)\citenamefont {Zhang}, \citenamefont {He}, \citenamefont {Eggert}, \citenamefont {Moessner},\ and\ \citenamefont {Pollmann}}]{zhang2018continuous}%
  \BibitemOpen
  \bibfield  {author} {\bibinfo {author} {\bibfnamefont {X.-F.}\ \bibnamefont {Zhang}}, \bibinfo {author} {\bibfnamefont {Y.-C.}\ \bibnamefont {He}}, \bibinfo {author} {\bibfnamefont {S.}~\bibnamefont {Eggert}}, \bibinfo {author} {\bibfnamefont {R.}~\bibnamefont {Moessner}},\ and\ \bibinfo {author} {\bibfnamefont {F.}~\bibnamefont {Pollmann}},\ }\bibfield  {title} {\bibinfo {title} {Continuous easy-plane deconfined phase transition on the kagome lattice},\ }\href {https://doi.org/10.1103/PhysRevLett.120.115702} {\bibfield  {journal} {\bibinfo  {journal} {Phys. Rev. Lett.}\ }\textbf {\bibinfo {volume} {120}},\ \bibinfo {pages} {115702} (\bibinfo {year} {2018})}\BibitemShut {NoStop}%
\bibitem [{\citenamefont {Liu}\ \emph {et~al.}(2024{\natexlab{b}})\citenamefont {Liu}, \citenamefont {Xiong}, \citenamefont {Xu},\ and\ \citenamefont {Zhang}}]{liu2024deconfined}%
  \BibitemOpen
  \bibfield  {author} {\bibinfo {author} {\bibfnamefont {D.-X.}\ \bibnamefont {Liu}}, \bibinfo {author} {\bibfnamefont {Z.}~\bibnamefont {Xiong}}, \bibinfo {author} {\bibfnamefont {Y.}~\bibnamefont {Xu}},\ and\ \bibinfo {author} {\bibfnamefont {X.-F.}\ \bibnamefont {Zhang}},\ }\bibfield  {title} {\bibinfo {title} {Deconfined quantum phase transition on the kagome lattice: Distinct velocities of spinon and string excitations},\ }\href {https://doi.org/10.1103/PhysRevB.109.L140404} {\bibfield  {journal} {\bibinfo  {journal} {Phys. Rev. B}\ }\textbf {\bibinfo {volume} {109}},\ \bibinfo {pages} {L140404} (\bibinfo {year} {2024}{\natexlab{b}})}\BibitemShut {NoStop}%
\bibitem [{\citenamefont {Liu}\ \emph {et~al.}(2022)\citenamefont {Liu}, \citenamefont {Hasik}, \citenamefont {Gong}, \citenamefont {Poilblanc}, \citenamefont {Chen},\ and\ \citenamefont {Gu}}]{liu2022emergence}%
  \BibitemOpen
  \bibfield  {author} {\bibinfo {author} {\bibfnamefont {W.-Y.}\ \bibnamefont {Liu}}, \bibinfo {author} {\bibfnamefont {J.}~\bibnamefont {Hasik}}, \bibinfo {author} {\bibfnamefont {S.-S.}\ \bibnamefont {Gong}}, \bibinfo {author} {\bibfnamefont {D.}~\bibnamefont {Poilblanc}}, \bibinfo {author} {\bibfnamefont {W.-Q.}\ \bibnamefont {Chen}},\ and\ \bibinfo {author} {\bibfnamefont {Z.-C.}\ \bibnamefont {Gu}},\ }\bibfield  {title} {\bibinfo {title} {Emergence of gapless quantum spin liquid from deconfined quantum critical point},\ }\href {https://doi.org/10.1103/PhysRevX.12.031039} {\bibfield  {journal} {\bibinfo  {journal} {Phys. Rev. X}\ }\textbf {\bibinfo {volume} {12}},\ \bibinfo {pages} {031039} (\bibinfo {year} {2022})}\BibitemShut {NoStop}%
\bibitem [{\citenamefont {Jim{\'e}nez}\ \emph {et~al.}(2021)\citenamefont {Jim{\'e}nez}, \citenamefont {Crone}, \citenamefont {Fogh}, \citenamefont {Zayed}, \citenamefont {Lortz}, \citenamefont {Pomjakushina}, \citenamefont {Conder}, \citenamefont {L{\"a}uchli}, \citenamefont {Weber}, \citenamefont {Wessel}, \citenamefont {Honecker}, \citenamefont {Normand}, \citenamefont {R{\"u}egg}, \citenamefont {Corboz}, \citenamefont {R{\o}nnow},\ and\ \citenamefont {Mila}}]{jimenezquantum2021}%
  \BibitemOpen
  \bibfield  {author} {\bibinfo {author} {\bibfnamefont {J.~L.}\ \bibnamefont {Jim{\'e}nez}}, \bibinfo {author} {\bibfnamefont {S.~P.~G.}\ \bibnamefont {Crone}}, \bibinfo {author} {\bibfnamefont {E.}~\bibnamefont {Fogh}}, \bibinfo {author} {\bibfnamefont {M.~E.}\ \bibnamefont {Zayed}}, \bibinfo {author} {\bibfnamefont {R.}~\bibnamefont {Lortz}}, \bibinfo {author} {\bibfnamefont {E.}~\bibnamefont {Pomjakushina}}, \bibinfo {author} {\bibfnamefont {K.}~\bibnamefont {Conder}}, \bibinfo {author} {\bibfnamefont {A.~M.}\ \bibnamefont {L{\"a}uchli}}, \bibinfo {author} {\bibfnamefont {L.}~\bibnamefont {Weber}}, \bibinfo {author} {\bibfnamefont {S.}~\bibnamefont {Wessel}}, \bibinfo {author} {\bibfnamefont {A.}~\bibnamefont {Honecker}}, \bibinfo {author} {\bibfnamefont {B.}~\bibnamefont {Normand}}, \bibinfo {author} {\bibfnamefont {C.}~\bibnamefont {R{\"u}egg}}, \bibinfo {author} {\bibfnamefont {P.}~\bibnamefont {Corboz}}, \bibinfo {author} {\bibfnamefont {H.~M.}\ \bibnamefont {R{\o}nnow}},\ and\ \bibinfo {author}
  {\bibfnamefont {F.}~\bibnamefont {Mila}},\ }\bibfield  {title} {\bibinfo {title} {A quantum magnetic analogue to the critical point of water},\ }\href {https://doi.org/10.1038/s41586-021-03411-8} {\bibfield  {journal} {\bibinfo  {journal} {Nature}\ }\textbf {\bibinfo {volume} {592}},\ \bibinfo {pages} {370} (\bibinfo {year} {2021})}\BibitemShut {NoStop}%
\bibitem [{\citenamefont {Zayed}\ \emph {et~al.}(2017)\citenamefont {Zayed}, \citenamefont {R{\"u}egg}, \citenamefont {Larrea~J.}, \citenamefont {L{\"a}uchli}, \citenamefont {Panagopoulos}, \citenamefont {Saxena}, \citenamefont {Ellerby}, \citenamefont {McMorrow}, \citenamefont {Str{\"a}ssle}, \citenamefont {Klotz}, \citenamefont {Hamel}, \citenamefont {Sadykov}, \citenamefont {Pomjakushin}, \citenamefont {Boehm}, \citenamefont {{Jim{\'e}nez{\textendash}Ruiz}}, \citenamefont {Schneidewind}, \citenamefont {Pomjakushina}, \citenamefont {Stingaciu}, \citenamefont {Conder},\ and\ \citenamefont {R{\o}nnow}}]{zayed4spin2017}%
  \BibitemOpen
  \bibfield  {author} {\bibinfo {author} {\bibfnamefont {M.~E.}\ \bibnamefont {Zayed}}, \bibinfo {author} {\bibfnamefont {C.}~\bibnamefont {R{\"u}egg}}, \bibinfo {author} {\bibfnamefont {J.}~\bibnamefont {Larrea~J.}}, \bibinfo {author} {\bibfnamefont {A.~M.}\ \bibnamefont {L{\"a}uchli}}, \bibinfo {author} {\bibfnamefont {C.}~\bibnamefont {Panagopoulos}}, \bibinfo {author} {\bibfnamefont {S.~S.}\ \bibnamefont {Saxena}}, \bibinfo {author} {\bibfnamefont {M.}~\bibnamefont {Ellerby}}, \bibinfo {author} {\bibfnamefont {D.~F.}\ \bibnamefont {McMorrow}}, \bibinfo {author} {\bibfnamefont {T.}~\bibnamefont {Str{\"a}ssle}}, \bibinfo {author} {\bibfnamefont {S.}~\bibnamefont {Klotz}}, \bibinfo {author} {\bibfnamefont {G.}~\bibnamefont {Hamel}}, \bibinfo {author} {\bibfnamefont {R.~A.}\ \bibnamefont {Sadykov}}, \bibinfo {author} {\bibfnamefont {V.}~\bibnamefont {Pomjakushin}}, \bibinfo {author} {\bibfnamefont {M.}~\bibnamefont {Boehm}}, \bibinfo {author} {\bibfnamefont {M.}~\bibnamefont {{Jim{\'e}nez{\textendash}Ruiz}}},
  \bibinfo {author} {\bibfnamefont {A.}~\bibnamefont {Schneidewind}}, \bibinfo {author} {\bibfnamefont {E.}~\bibnamefont {Pomjakushina}}, \bibinfo {author} {\bibfnamefont {M.}~\bibnamefont {Stingaciu}}, \bibinfo {author} {\bibfnamefont {K.}~\bibnamefont {Conder}},\ and\ \bibinfo {author} {\bibfnamefont {H.~M.}\ \bibnamefont {R{\o}nnow}},\ }\bibfield  {title} {\bibinfo {title} {4-spin plaquette singlet state in the {{Shastry}}\textendash{{Sutherland}} compound {{SrCu$_2$}}({{BO$_3$}})$_2$},\ }\href {https://doi.org/10.1038/nphys4190} {\bibfield  {journal} {\bibinfo  {journal} {Nat. Phys.}\ }\textbf {\bibinfo {volume} {13}},\ \bibinfo {pages} {962} (\bibinfo {year} {2017})}\BibitemShut {NoStop}%
\bibitem [{\citenamefont {Guo}\ \emph {et~al.}(2020)\citenamefont {Guo}, \citenamefont {Sun}, \citenamefont {Zhao}, \citenamefont {Wang}, \citenamefont {Hong}, \citenamefont {Sidorov}, \citenamefont {Ma}, \citenamefont {Wu}, \citenamefont {Li}, \citenamefont {Meng}, \citenamefont {Sandvik},\ and\ \citenamefont {Sun}}]{guoQuantum2020}%
  \BibitemOpen
  \bibfield  {author} {\bibinfo {author} {\bibfnamefont {J.}~\bibnamefont {Guo}}, \bibinfo {author} {\bibfnamefont {G.}~\bibnamefont {Sun}}, \bibinfo {author} {\bibfnamefont {B.}~\bibnamefont {Zhao}}, \bibinfo {author} {\bibfnamefont {L.}~\bibnamefont {Wang}}, \bibinfo {author} {\bibfnamefont {W.}~\bibnamefont {Hong}}, \bibinfo {author} {\bibfnamefont {V.~A.}\ \bibnamefont {Sidorov}}, \bibinfo {author} {\bibfnamefont {N.}~\bibnamefont {Ma}}, \bibinfo {author} {\bibfnamefont {Q.}~\bibnamefont {Wu}}, \bibinfo {author} {\bibfnamefont {S.}~\bibnamefont {Li}}, \bibinfo {author} {\bibfnamefont {Z.~Y.}\ \bibnamefont {Meng}}, \bibinfo {author} {\bibfnamefont {A.~W.}\ \bibnamefont {Sandvik}},\ and\ \bibinfo {author} {\bibfnamefont {L.}~\bibnamefont {Sun}},\ }\bibfield  {title} {\bibinfo {title} {Quantum phases of ${\mathrm{srcu}}_{2}({\mathrm{bo}}_{3}{)}_{2}$ from high-pressure thermodynamics},\ }\href {https://doi.org/10.1103/PhysRevLett.124.206602} {\bibfield  {journal} {\bibinfo  {journal} {Phys. Rev. Lett.}\
  }\textbf {\bibinfo {volume} {124}},\ \bibinfo {pages} {206602} (\bibinfo {year} {2020})}\BibitemShut {NoStop}%
\bibitem [{\citenamefont {Sun}\ \emph {et~al.}(2021{\natexlab{b}})\citenamefont {Sun}, \citenamefont {Ma}, \citenamefont {Zhao}, \citenamefont {Sandvik},\ and\ \citenamefont {Meng}}]{sunEmergent2021}%
  \BibitemOpen
  \bibfield  {author} {\bibinfo {author} {\bibfnamefont {G.}~\bibnamefont {Sun}}, \bibinfo {author} {\bibfnamefont {N.}~\bibnamefont {Ma}}, \bibinfo {author} {\bibfnamefont {B.}~\bibnamefont {Zhao}}, \bibinfo {author} {\bibfnamefont {A.~W.}\ \bibnamefont {Sandvik}},\ and\ \bibinfo {author} {\bibfnamefont {Z.~Y.}\ \bibnamefont {Meng}},\ }\bibfield  {title} {\bibinfo {title} {Emergent o(4) symmetry at the phase transition from plaquette-singlet to antiferromagnetic order in quasi-two-dimensional quantum magnets},\ }\href {https://doi.org/10.1088/1674-1056/abf3b8} {\bibfield  {journal} {\bibinfo  {journal} {Chin. Phys. B}\ }\textbf {\bibinfo {volume} {30}},\ \bibinfo {eid} {067505} (\bibinfo {year} {2021}{\natexlab{b}})}\BibitemShut {NoStop}%
\bibitem [{\citenamefont {Cui}\ \emph {et~al.}(2023)\citenamefont {Cui}, \citenamefont {Liu}, \citenamefont {Lin}, \citenamefont {Wu}, \citenamefont {Hong}, \citenamefont {Liu}, \citenamefont {Li}, \citenamefont {Hu}, \citenamefont {Xi}, \citenamefont {Li}, \citenamefont {Yu}, \citenamefont {Sandvik},\ and\ \citenamefont {Yu}}]{cuiProximate2023}%
  \BibitemOpen
  \bibfield  {author} {\bibinfo {author} {\bibfnamefont {Y.}~\bibnamefont {Cui}}, \bibinfo {author} {\bibfnamefont {L.}~\bibnamefont {Liu}}, \bibinfo {author} {\bibfnamefont {H.}~\bibnamefont {Lin}}, \bibinfo {author} {\bibfnamefont {K.-H.}\ \bibnamefont {Wu}}, \bibinfo {author} {\bibfnamefont {W.}~\bibnamefont {Hong}}, \bibinfo {author} {\bibfnamefont {X.}~\bibnamefont {Liu}}, \bibinfo {author} {\bibfnamefont {C.}~\bibnamefont {Li}}, \bibinfo {author} {\bibfnamefont {Z.}~\bibnamefont {Hu}}, \bibinfo {author} {\bibfnamefont {N.}~\bibnamefont {Xi}}, \bibinfo {author} {\bibfnamefont {S.}~\bibnamefont {Li}}, \bibinfo {author} {\bibfnamefont {R.}~\bibnamefont {Yu}}, \bibinfo {author} {\bibfnamefont {A.~W.}\ \bibnamefont {Sandvik}},\ and\ \bibinfo {author} {\bibfnamefont {W.}~\bibnamefont {Yu}},\ }\bibfield  {title} {\bibinfo {title} {Proximate deconfined quantum critical point in srcu$_2$(bo$_3$)$_2$},\ }\href {https://doi.org/10.1126/science.adc9487} {\bibfield  {journal} {\bibinfo  {journal} {Science}\ }\textbf
  {\bibinfo {volume} {380}},\ \bibinfo {pages} {1179} (\bibinfo {year} {2023})}\BibitemShut {NoStop}%
\bibitem [{\citenamefont {Guo}\ \emph {et~al.}(2025)\citenamefont {Guo}, \citenamefont {Wang}, \citenamefont {Huang}, \citenamefont {Chen}, \citenamefont {Hong}, \citenamefont {Cai}, \citenamefont {Zhao}, \citenamefont {Han}, \citenamefont {Chen}, \citenamefont {Zhou}, \citenamefont {Li}, \citenamefont {Wu}, \citenamefont {Meng},\ and\ \citenamefont {Sun}}]{guoDeconfined2023}%
  \BibitemOpen
  \bibfield  {author} {\bibinfo {author} {\bibfnamefont {J.}~\bibnamefont {Guo}}, \bibinfo {author} {\bibfnamefont {P.}~\bibnamefont {Wang}}, \bibinfo {author} {\bibfnamefont {C.}~\bibnamefont {Huang}}, \bibinfo {author} {\bibfnamefont {B.-B.}\ \bibnamefont {Chen}}, \bibinfo {author} {\bibfnamefont {W.}~\bibnamefont {Hong}}, \bibinfo {author} {\bibfnamefont {S.}~\bibnamefont {Cai}}, \bibinfo {author} {\bibfnamefont {J.}~\bibnamefont {Zhao}}, \bibinfo {author} {\bibfnamefont {J.}~\bibnamefont {Han}}, \bibinfo {author} {\bibfnamefont {X.}~\bibnamefont {Chen}}, \bibinfo {author} {\bibfnamefont {Y.}~\bibnamefont {Zhou}}, \bibinfo {author} {\bibfnamefont {S.}~\bibnamefont {Li}}, \bibinfo {author} {\bibfnamefont {Q.}~\bibnamefont {Wu}}, \bibinfo {author} {\bibfnamefont {Z.~Y.}\ \bibnamefont {Meng}},\ and\ \bibinfo {author} {\bibfnamefont {L.}~\bibnamefont {Sun}},\ }\bibfield  {title} {\bibinfo {title} {Deconfined quantum critical point lost in pressurized srcu2(bo3)2},\ }\href
  {https://doi.org/10.1038/s42005-025-01976-8} {\bibfield  {journal} {\bibinfo  {journal} {Communications Physics}\ }\textbf {\bibinfo {volume} {8}},\ \bibinfo {pages} {75} (\bibinfo {year} {2025})}\BibitemShut {NoStop}%
\bibitem [{\citenamefont {Cui}\ \emph {et~al.}(2025)\citenamefont {Cui}, \citenamefont {Yu},\ and\ \citenamefont {Yu}}]{cui2025deconfined}%
  \BibitemOpen
  \bibfield  {author} {\bibinfo {author} {\bibfnamefont {Y.}~\bibnamefont {Cui}}, \bibinfo {author} {\bibfnamefont {R.}~\bibnamefont {Yu}},\ and\ \bibinfo {author} {\bibfnamefont {W.}~\bibnamefont {Yu}},\ }\bibfield  {title} {\bibinfo {title} {Deconfined quantum critical point: A review of progress},\ }\href {https://doi.org/10.1088/0256-307X/42/4/047503} {\bibfield  {journal} {\bibinfo  {journal} {Chinese Physics Letters}\ }\textbf {\bibinfo {volume} {42}},\ \bibinfo {pages} {047503} (\bibinfo {year} {2025})}\BibitemShut {NoStop}%
\bibitem [{\citenamefont {Takahashi}\ \emph {et~al.}(2024)\citenamefont {Takahashi}, \citenamefont {Shao}, \citenamefont {Zhao}, \citenamefont {Guo},\ and\ \citenamefont {Sandvik}}]{takahashi2024so}%
  \BibitemOpen
  \bibfield  {author} {\bibinfo {author} {\bibfnamefont {J.}~\bibnamefont {Takahashi}}, \bibinfo {author} {\bibfnamefont {H.}~\bibnamefont {Shao}}, \bibinfo {author} {\bibfnamefont {B.}~\bibnamefont {Zhao}}, \bibinfo {author} {\bibfnamefont {W.}~\bibnamefont {Guo}},\ and\ \bibinfo {author} {\bibfnamefont {A.~W.}\ \bibnamefont {Sandvik}},\ }\bibfield  {title} {\bibinfo {title} {So (5) multicriticality in two-dimensional quantum magnets},\ }\href@noop {} {\bibfield  {journal} {\bibinfo  {journal} {arXiv preprint arXiv:2405.06607}\ } (\bibinfo {year} {2024})}\BibitemShut {NoStop}%
\bibitem [{\citenamefont {Zhao}\ \emph {et~al.}(2022)\citenamefont {Zhao}, \citenamefont {Wang}, \citenamefont {Yan}, \citenamefont {Cheng},\ and\ \citenamefont {Meng}}]{zhaoScaling2022}%
  \BibitemOpen
  \bibfield  {author} {\bibinfo {author} {\bibfnamefont {J.}~\bibnamefont {Zhao}}, \bibinfo {author} {\bibfnamefont {Y.-C.}\ \bibnamefont {Wang}}, \bibinfo {author} {\bibfnamefont {Z.}~\bibnamefont {Yan}}, \bibinfo {author} {\bibfnamefont {M.}~\bibnamefont {Cheng}},\ and\ \bibinfo {author} {\bibfnamefont {Z.~Y.}\ \bibnamefont {Meng}},\ }\bibfield  {title} {\bibinfo {title} {Scaling of entanglement entropy at deconfined quantum criticality},\ }\href {https://doi.org/10.1103/PhysRevLett.128.010601} {\bibfield  {journal} {\bibinfo  {journal} {Phys. Rev. Lett.}\ }\textbf {\bibinfo {volume} {128}},\ \bibinfo {pages} {010601} (\bibinfo {year} {2022})}\BibitemShut {NoStop}%
\bibitem [{\citenamefont {Deng}\ \emph {et~al.}(2024)\citenamefont {Deng}, \citenamefont {Liu}, \citenamefont {Guo},\ and\ \citenamefont {Lin}}]{deng2024diagnosing}%
  \BibitemOpen
  \bibfield  {author} {\bibinfo {author} {\bibfnamefont {Z.}~\bibnamefont {Deng}}, \bibinfo {author} {\bibfnamefont {L.}~\bibnamefont {Liu}}, \bibinfo {author} {\bibfnamefont {W.}~\bibnamefont {Guo}},\ and\ \bibinfo {author} {\bibfnamefont {H.-Q.}\ \bibnamefont {Lin}},\ }\bibfield  {title} {\bibinfo {title} {Diagnosing quantum phase transition order and deconfined criticality via entanglement entropy},\ }\href {https://doi.org/10.1103/PhysRevLett.133.100402} {\bibfield  {journal} {\bibinfo  {journal} {Phys. Rev. Lett.}\ }\textbf {\bibinfo {volume} {133}},\ \bibinfo {pages} {100402} (\bibinfo {year} {2024})}\BibitemShut {NoStop}%
\bibitem [{\citenamefont {Song}\ \emph {et~al.}(2025)\citenamefont {Song}, \citenamefont {Zhao}, \citenamefont {Cheng}, \citenamefont {Xu}, \citenamefont {Scherer}, \citenamefont {Janssen},\ and\ \citenamefont {Meng}}]{song2023deconfined}%
  \BibitemOpen
  \bibfield  {author} {\bibinfo {author} {\bibfnamefont {M.}~\bibnamefont {Song}}, \bibinfo {author} {\bibfnamefont {J.}~\bibnamefont {Zhao}}, \bibinfo {author} {\bibfnamefont {M.}~\bibnamefont {Cheng}}, \bibinfo {author} {\bibfnamefont {C.}~\bibnamefont {Xu}}, \bibinfo {author} {\bibfnamefont {M.}~\bibnamefont {Scherer}}, \bibinfo {author} {\bibfnamefont {L.}~\bibnamefont {Janssen}},\ and\ \bibinfo {author} {\bibfnamefont {Z.~Y.}\ \bibnamefont {Meng}},\ }\bibfield  {title} {\bibinfo {title} {Evolution of entanglement entropy at su(<i>n</i>) deconfined quantum critical points},\ }\href {https://doi.org/10.1126/sciadv.adr0634} {\bibfield  {journal} {\bibinfo  {journal} {Science Advances}\ }\textbf {\bibinfo {volume} {11}},\ \bibinfo {pages} {eadr0634} (\bibinfo {year} {2025})}\BibitemShut {NoStop}%
\bibitem [{\citenamefont {Zhou}\ \emph {et~al.}(2024)\citenamefont {Zhou}, \citenamefont {Hu}, \citenamefont {Zhu},\ and\ \citenamefont {He}}]{zhou2024mathrmso5}%
  \BibitemOpen
  \bibfield  {author} {\bibinfo {author} {\bibfnamefont {Z.}~\bibnamefont {Zhou}}, \bibinfo {author} {\bibfnamefont {L.}~\bibnamefont {Hu}}, \bibinfo {author} {\bibfnamefont {W.}~\bibnamefont {Zhu}},\ and\ \bibinfo {author} {\bibfnamefont {Y.-C.}\ \bibnamefont {He}},\ }\bibfield  {title} {\bibinfo {title} {So(5) deconfined phase transition under the fuzzy-sphere microscope: Approximate conformal symmetry, pseudo-criticality, and operator spectrum},\ }\href {https://doi.org/10.1103/PhysRevX.14.021044} {\bibfield  {journal} {\bibinfo  {journal} {Phys. Rev. X}\ }\textbf {\bibinfo {volume} {14}},\ \bibinfo {pages} {021044} (\bibinfo {year} {2024})}\BibitemShut {NoStop}%
\bibitem [{\citenamefont {Anderson}(1952)}]{anderson1952approximate}%
  \BibitemOpen
  \bibfield  {author} {\bibinfo {author} {\bibfnamefont {P.~W.}\ \bibnamefont {Anderson}},\ }\bibfield  {title} {\bibinfo {title} {An approximate quantum theory of the antiferromagnetic ground state},\ }\href@noop {} {\bibfield  {journal} {\bibinfo  {journal} {Physical Review}\ }\textbf {\bibinfo {volume} {86}},\ \bibinfo {pages} {694} (\bibinfo {year} {1952})}\BibitemShut {NoStop}%
\bibitem [{\citenamefont {Lhuillier}(2005)}]{lhuillier2005frustratedquantummagnets}%
  \BibitemOpen
  \bibfield  {author} {\bibinfo {author} {\bibfnamefont {C.}~\bibnamefont {Lhuillier}},\ }\href {https://arxiv.org/abs/cond-mat/0502464} {\bibinfo {title} {Frustrated quantum magnets}} (\bibinfo {year} {2005}),\ \Eprint {https://arxiv.org/abs/cond-mat/0502464} {arXiv:cond-mat/0502464} \BibitemShut {NoStop}%
\bibitem [{\citenamefont {Wietek}\ \emph {et~al.}(2017)\citenamefont {Wietek}, \citenamefont {Schuler},\ and\ \citenamefont {Läuchli}}]{wietek2017studying}%
  \BibitemOpen
  \bibfield  {author} {\bibinfo {author} {\bibfnamefont {A.}~\bibnamefont {Wietek}}, \bibinfo {author} {\bibfnamefont {M.}~\bibnamefont {Schuler}},\ and\ \bibinfo {author} {\bibfnamefont {A.~M.}\ \bibnamefont {Läuchli}},\ }\href {https://arxiv.org/abs/1704.08622} {\bibinfo {title} {Studying continuous symmetry breaking using energy level spectroscopy}} (\bibinfo {year} {2017}),\ \Eprint {https://arxiv.org/abs/1704.08622} {arXiv:1704.08622} \BibitemShut {NoStop}%
\bibitem [{Note1()}]{Note1}%
  \BibitemOpen
  \bibinfo {note} {A CFT tower is formed if conformal field theory is valid for critical points; for the (2+1)D $J$-$Q$ model of a DQCP, this remains controversial}\BibitemShut {NoStop}%
\bibitem [{\citenamefont {Alba}\ \emph {et~al.}(2013)\citenamefont {Alba}, \citenamefont {Haque},\ and\ \citenamefont {L\"auchli}}]{Alba2013entanglement}%
  \BibitemOpen
  \bibfield  {author} {\bibinfo {author} {\bibfnamefont {V.}~\bibnamefont {Alba}}, \bibinfo {author} {\bibfnamefont {M.}~\bibnamefont {Haque}},\ and\ \bibinfo {author} {\bibfnamefont {A.~M.}\ \bibnamefont {L\"auchli}},\ }\bibfield  {title} {\bibinfo {title} {Entanglement spectrum of the two-dimensional bose-hubbard model},\ }\href {https://doi.org/10.1103/PhysRevLett.110.260403} {\bibfield  {journal} {\bibinfo  {journal} {Phys. Rev. Lett.}\ }\textbf {\bibinfo {volume} {110}},\ \bibinfo {pages} {260403} (\bibinfo {year} {2013})}\BibitemShut {NoStop}%
\bibitem [{\citenamefont {Kolley}\ \emph {et~al.}(2013)\citenamefont {Kolley}, \citenamefont {Depenbrock}, \citenamefont {McCulloch}, \citenamefont {Schollw\"ock},\ and\ \citenamefont {Alba}}]{Kolley2013entanglement}%
  \BibitemOpen
  \bibfield  {author} {\bibinfo {author} {\bibfnamefont {F.}~\bibnamefont {Kolley}}, \bibinfo {author} {\bibfnamefont {S.}~\bibnamefont {Depenbrock}}, \bibinfo {author} {\bibfnamefont {I.~P.}\ \bibnamefont {McCulloch}}, \bibinfo {author} {\bibfnamefont {U.}~\bibnamefont {Schollw\"ock}},\ and\ \bibinfo {author} {\bibfnamefont {V.}~\bibnamefont {Alba}},\ }\bibfield  {title} {\bibinfo {title} {Entanglement spectroscopy of su(2)-broken phases in two dimensions},\ }\href {https://doi.org/10.1103/PhysRevB.88.144426} {\bibfield  {journal} {\bibinfo  {journal} {Phys. Rev. B}\ }\textbf {\bibinfo {volume} {88}},\ \bibinfo {pages} {144426} (\bibinfo {year} {2013})}\BibitemShut {NoStop}%
\bibitem [{\citenamefont {Mao}\ \emph {et~al.}(2025)\citenamefont {Mao}, \citenamefont {Ding}, \citenamefont {Wang}, \citenamefont {Hu},\ and\ \citenamefont {Yan}}]{mao2025sampling}%
  \BibitemOpen
  \bibfield  {author} {\bibinfo {author} {\bibfnamefont {B.-B.}\ \bibnamefont {Mao}}, \bibinfo {author} {\bibfnamefont {Y.-M.}\ \bibnamefont {Ding}}, \bibinfo {author} {\bibfnamefont {Z.}~\bibnamefont {Wang}}, \bibinfo {author} {\bibfnamefont {S.}~\bibnamefont {Hu}},\ and\ \bibinfo {author} {\bibfnamefont {Z.}~\bibnamefont {Yan}},\ }\bibfield  {title} {\bibinfo {title} {Sampling reduced density matrix to extract fine levels of entanglement spectrum and restore entanglement hamiltonian},\ }\href {https://doi.org/10.1038/s41467-025-58058-0} {\bibfield  {journal} {\bibinfo  {journal} {Nature Communications}\ }\textbf {\bibinfo {volume} {16}},\ \bibinfo {pages} {2880} (\bibinfo {year} {2025})}\BibitemShut {NoStop}%
\bibitem [{\citenamefont {Rams}\ \emph {et~al.}(2018)\citenamefont {Rams}, \citenamefont {Czarnik},\ and\ \citenamefont {Cincio}}]{rams2018precise}%
  \BibitemOpen
  \bibfield  {author} {\bibinfo {author} {\bibfnamefont {M.~M.}\ \bibnamefont {Rams}}, \bibinfo {author} {\bibfnamefont {P.}~\bibnamefont {Czarnik}},\ and\ \bibinfo {author} {\bibfnamefont {L.}~\bibnamefont {Cincio}},\ }\bibfield  {title} {\bibinfo {title} {Precise extrapolation of the correlation function asymptotics in uniform tensor network states with application to the bose-hubbard and xxz models},\ }\href {https://doi.org/10.1103/PhysRevX.8.041033} {\bibfield  {journal} {\bibinfo  {journal} {Phys. Rev. X}\ }\textbf {\bibinfo {volume} {8}},\ \bibinfo {pages} {041033} (\bibinfo {year} {2018})}\BibitemShut {NoStop}%
\bibitem [{\citenamefont {Chandran}\ \emph {et~al.}(2014)\citenamefont {Chandran}, \citenamefont {Khemani},\ and\ \citenamefont {Sondhi}}]{chandran2013how}%
  \BibitemOpen
  \bibfield  {author} {\bibinfo {author} {\bibfnamefont {A.}~\bibnamefont {Chandran}}, \bibinfo {author} {\bibfnamefont {V.}~\bibnamefont {Khemani}},\ and\ \bibinfo {author} {\bibfnamefont {S.~L.}\ \bibnamefont {Sondhi}},\ }\bibfield  {title} {\bibinfo {title} {How universal is the entanglement spectrum?},\ }\href {https://doi.org/10.1103/PhysRevLett.113.060501} {\bibfield  {journal} {\bibinfo  {journal} {Phys. Rev. Lett.}\ }\textbf {\bibinfo {volume} {113}},\ \bibinfo {pages} {060501} (\bibinfo {year} {2014})}\BibitemShut {NoStop}%
\bibitem [{\citenamefont {Poilblanc}(2010)}]{Poilblanc2010entanglement}%
  \BibitemOpen
  \bibfield  {author} {\bibinfo {author} {\bibfnamefont {D.}~\bibnamefont {Poilblanc}},\ }\bibfield  {title} {\bibinfo {title} {Entanglement spectra of quantum heisenberg ladders},\ }\href {https://doi.org/10.1103/PhysRevLett.105.077202} {\bibfield  {journal} {\bibinfo  {journal} {Phys. Rev. Lett.}\ }\textbf {\bibinfo {volume} {105}},\ \bibinfo {pages} {077202} (\bibinfo {year} {2010})}\BibitemShut {NoStop}%
\bibitem [{\citenamefont {Yan}\ and\ \citenamefont {Meng}(2023)}]{zyan2021entanglement}%
  \BibitemOpen
  \bibfield  {author} {\bibinfo {author} {\bibfnamefont {Z.}~\bibnamefont {Yan}}\ and\ \bibinfo {author} {\bibfnamefont {Z.~Y.}\ \bibnamefont {Meng}},\ }\bibfield  {title} {\bibinfo {title} {Unlocking the general relationship between energy and entanglement spectra via the wormhole effect},\ }\href {https://doi.org/10.1038/s41467-023-37756-7} {\bibfield  {journal} {\bibinfo  {journal} {Nature Communications}\ }\textbf {\bibinfo {volume} {14}},\ \bibinfo {pages} {2360} (\bibinfo {year} {2023})}\BibitemShut {NoStop}%
\bibitem [{\citenamefont {Song}\ \emph {et~al.}(2023)\citenamefont {Song}, \citenamefont {Zhao}, \citenamefont {Yan},\ and\ \citenamefont {Meng}}]{song2023different}%
  \BibitemOpen
  \bibfield  {author} {\bibinfo {author} {\bibfnamefont {M.}~\bibnamefont {Song}}, \bibinfo {author} {\bibfnamefont {J.}~\bibnamefont {Zhao}}, \bibinfo {author} {\bibfnamefont {Z.}~\bibnamefont {Yan}},\ and\ \bibinfo {author} {\bibfnamefont {Z.~Y.}\ \bibnamefont {Meng}},\ }\bibfield  {title} {\bibinfo {title} {Different temperature dependence for the edge and bulk of the entanglement hamiltonian},\ }\href {https://doi.org/10.1103/PhysRevB.108.075114} {\bibfield  {journal} {\bibinfo  {journal} {Phys. Rev. B}\ }\textbf {\bibinfo {volume} {108}},\ \bibinfo {pages} {075114} (\bibinfo {year} {2023})}\BibitemShut {NoStop}%
\bibitem [{\citenamefont {Li}\ \emph {et~al.}(2024)\citenamefont {Li}, \citenamefont {Huang}, \citenamefont {Ding}, \citenamefont {Meng}, \citenamefont {Wang},\ and\ \citenamefont {Yan}}]{li2023relevant}%
  \BibitemOpen
  \bibfield  {author} {\bibinfo {author} {\bibfnamefont {C.}~\bibnamefont {Li}}, \bibinfo {author} {\bibfnamefont {R.-Z.}\ \bibnamefont {Huang}}, \bibinfo {author} {\bibfnamefont {Y.-M.}\ \bibnamefont {Ding}}, \bibinfo {author} {\bibfnamefont {Z.~Y.}\ \bibnamefont {Meng}}, \bibinfo {author} {\bibfnamefont {Y.-C.}\ \bibnamefont {Wang}},\ and\ \bibinfo {author} {\bibfnamefont {Z.}~\bibnamefont {Yan}},\ }\bibfield  {title} {\bibinfo {title} {Relevant long-range interaction of the entanglement hamiltonian emerges from a short-range gapped system},\ }\href {https://doi.org/10.1103/PhysRevB.109.195169} {\bibfield  {journal} {\bibinfo  {journal} {Phys. Rev. B}\ }\textbf {\bibinfo {volume} {109}},\ \bibinfo {pages} {195169} (\bibinfo {year} {2024})}\BibitemShut {NoStop}%
\bibitem [{\citenamefont {Wu}\ \emph {et~al.}(2023)\citenamefont {Wu}, \citenamefont {Ran}, \citenamefont {Yin}, \citenamefont {Li}, \citenamefont {Mao}, \citenamefont {Wang},\ and\ \citenamefont {Yan}}]{wu2023classical}%
  \BibitemOpen
  \bibfield  {author} {\bibinfo {author} {\bibfnamefont {S.}~\bibnamefont {Wu}}, \bibinfo {author} {\bibfnamefont {X.}~\bibnamefont {Ran}}, \bibinfo {author} {\bibfnamefont {B.}~\bibnamefont {Yin}}, \bibinfo {author} {\bibfnamefont {Q.-F.}\ \bibnamefont {Li}}, \bibinfo {author} {\bibfnamefont {B.-B.}\ \bibnamefont {Mao}}, \bibinfo {author} {\bibfnamefont {Y.-C.}\ \bibnamefont {Wang}},\ and\ \bibinfo {author} {\bibfnamefont {Z.}~\bibnamefont {Yan}},\ }\bibfield  {title} {\bibinfo {title} {Classical model emerges in quantum entanglement: Quantum monte carlo study for an ising-heisenberg bilayer},\ }\href {https://doi.org/10.1103/PhysRevB.107.155121} {\bibfield  {journal} {\bibinfo  {journal} {Phys. Rev. B}\ }\textbf {\bibinfo {volume} {107}},\ \bibinfo {pages} {155121} (\bibinfo {year} {2023})}\BibitemShut {NoStop}%
\bibitem [{\citenamefont {Li}\ and\ \citenamefont {Haldane}(2008)}]{Li2008entangle}%
  \BibitemOpen
  \bibfield  {author} {\bibinfo {author} {\bibfnamefont {H.}~\bibnamefont {Li}}\ and\ \bibinfo {author} {\bibfnamefont {F.~D.~M.}\ \bibnamefont {Haldane}},\ }\bibfield  {title} {\bibinfo {title} {Entanglement spectrum as a generalization of entanglement entropy: Identification of topological order in non-abelian fractional quantum hall effect states},\ }\href {https://doi.org/10.1103/PhysRevLett.101.010504} {\bibfield  {journal} {\bibinfo  {journal} {Phys. Rev. Lett.}\ }\textbf {\bibinfo {volume} {101}},\ \bibinfo {pages} {010504} (\bibinfo {year} {2008})}\BibitemShut {NoStop}%
\bibitem [{\citenamefont {Qi}\ \emph {et~al.}(2012)\citenamefont {Qi}, \citenamefont {Katsura},\ and\ \citenamefont {Ludwig}}]{XLQi2012}%
  \BibitemOpen
  \bibfield  {author} {\bibinfo {author} {\bibfnamefont {X.-L.}\ \bibnamefont {Qi}}, \bibinfo {author} {\bibfnamefont {H.}~\bibnamefont {Katsura}},\ and\ \bibinfo {author} {\bibfnamefont {A.~W.~W.}\ \bibnamefont {Ludwig}},\ }\bibfield  {title} {\bibinfo {title} {General relationship between the entanglement spectrum and the edge state spectrum of topological quantum states},\ }\href {https://doi.org/10.1103/PhysRevLett.108.196402} {\bibfield  {journal} {\bibinfo  {journal} {Phys. Rev. Lett.}\ }\textbf {\bibinfo {volume} {108}},\ \bibinfo {pages} {196402} (\bibinfo {year} {2012})}\BibitemShut {NoStop}%
\bibitem [{Note2()}]{Note2}%
  \BibitemOpen
  \bibinfo {note} {Noting that the phase transition line can also be lower than $T_E=1$ if the phase at $T_E=1$ is disordered.}\BibitemShut {Stop}%
\bibitem [{\citenamefont {Ma}\ \emph {et~al.}(2018{\natexlab{b}})\citenamefont {Ma}, \citenamefont {Weinberg}, \citenamefont {Shao}, \citenamefont {Guo}, \citenamefont {Yao},\ and\ \citenamefont {Sandvik}}]{NSMa2018anomalous}%
  \BibitemOpen
  \bibfield  {author} {\bibinfo {author} {\bibfnamefont {N.}~\bibnamefont {Ma}}, \bibinfo {author} {\bibfnamefont {P.}~\bibnamefont {Weinberg}}, \bibinfo {author} {\bibfnamefont {H.}~\bibnamefont {Shao}}, \bibinfo {author} {\bibfnamefont {W.}~\bibnamefont {Guo}}, \bibinfo {author} {\bibfnamefont {D.-X.}\ \bibnamefont {Yao}},\ and\ \bibinfo {author} {\bibfnamefont {A.~W.}\ \bibnamefont {Sandvik}},\ }\bibfield  {title} {\bibinfo {title} {Anomalous quantum-critical scaling corrections in two-dimensional antiferromagnets},\ }\href {https://doi.org/10.1103/PhysRevLett.121.117202} {\bibfield  {journal} {\bibinfo  {journal} {Phys. Rev. Lett.}\ }\textbf {\bibinfo {volume} {121}},\ \bibinfo {pages} {117202} (\bibinfo {year} {2018}{\natexlab{b}})}\BibitemShut {NoStop}%
\bibitem [{\citenamefont {Matsumoto}\ \emph {et~al.}(2001)\citenamefont {Matsumoto}, \citenamefont {Yasuda}, \citenamefont {Todo},\ and\ \citenamefont {Takayama}}]{Matsumoto2001}%
  \BibitemOpen
  \bibfield  {author} {\bibinfo {author} {\bibfnamefont {M.}~\bibnamefont {Matsumoto}}, \bibinfo {author} {\bibfnamefont {C.}~\bibnamefont {Yasuda}}, \bibinfo {author} {\bibfnamefont {S.}~\bibnamefont {Todo}},\ and\ \bibinfo {author} {\bibfnamefont {H.}~\bibnamefont {Takayama}},\ }\bibfield  {title} {\bibinfo {title} {Ground-state phase diagram of quantum heisenberg antiferromagnets on the anisotropic dimerized square lattice},\ }\href {https://doi.org/10.1103/PhysRevB.65.014407} {\bibfield  {journal} {\bibinfo  {journal} {Phys. Rev. B}\ }\textbf {\bibinfo {volume} {65}},\ \bibinfo {pages} {014407} (\bibinfo {year} {2001})}\BibitemShut {NoStop}%
\bibitem [{\citenamefont {Metlitski}\ and\ \citenamefont {Grover}(2015)}]{metlitski2015entanglement}%
  \BibitemOpen
  \bibfield  {author} {\bibinfo {author} {\bibfnamefont {M.~A.}\ \bibnamefont {Metlitski}}\ and\ \bibinfo {author} {\bibfnamefont {T.}~\bibnamefont {Grover}},\ }\href@noop {} {\bibinfo {title} {Entanglement entropy of systems with spontaneously broken continuous symmetry}} (\bibinfo {year} {2015}),\ \Eprint {https://arxiv.org/abs/1112.5166} {arXiv:1112.5166} \BibitemShut {NoStop}%
\bibitem [{\citenamefont {Penc}\ \emph {et~al.}(2003)\citenamefont {Penc}, \citenamefont {Mambrini}, \citenamefont {Fazekas},\ and\ \citenamefont {Mila}}]{Penc2003PRBSU4}%
  \BibitemOpen
  \bibfield  {author} {\bibinfo {author} {\bibfnamefont {K.}~\bibnamefont {Penc}}, \bibinfo {author} {\bibfnamefont {M.}~\bibnamefont {Mambrini}}, \bibinfo {author} {\bibfnamefont {P.}~\bibnamefont {Fazekas}},\ and\ \bibinfo {author} {\bibfnamefont {F.}~\bibnamefont {Mila}},\ }\bibfield  {title} {\bibinfo {title} {Quantum phase transition in the su(4) spin-orbital model on the triangular lattice},\ }\href {https://doi.org/10.1103/PhysRevB.68.012408} {\bibfield  {journal} {\bibinfo  {journal} {Phys. Rev. B}\ }\textbf {\bibinfo {volume} {68}},\ \bibinfo {pages} {012408} (\bibinfo {year} {2003})}\BibitemShut {NoStop}%
\bibitem [{\citenamefont {T\'oth}\ \emph {et~al.}(2010)\citenamefont {T\'oth}, \citenamefont {L\"auchli}, \citenamefont {Mila},\ and\ \citenamefont {Penc}}]{Toth2010PRLSU3}%
  \BibitemOpen
  \bibfield  {author} {\bibinfo {author} {\bibfnamefont {T.~A.}\ \bibnamefont {T\'oth}}, \bibinfo {author} {\bibfnamefont {A.~M.}\ \bibnamefont {L\"auchli}}, \bibinfo {author} {\bibfnamefont {F.}~\bibnamefont {Mila}},\ and\ \bibinfo {author} {\bibfnamefont {K.}~\bibnamefont {Penc}},\ }\bibfield  {title} {\bibinfo {title} {Three-sublattice ordering of the su(3) heisenberg model of three-flavor fermions on the square and cubic lattices},\ }\href {https://doi.org/10.1103/PhysRevLett.105.265301} {\bibfield  {journal} {\bibinfo  {journal} {Phys. Rev. Lett.}\ }\textbf {\bibinfo {volume} {105}},\ \bibinfo {pages} {265301} (\bibinfo {year} {2010})}\BibitemShut {NoStop}%
\bibitem [{\citenamefont {Corboz}\ \emph {et~al.}(2011)\citenamefont {Corboz}, \citenamefont {L\"auchli}, \citenamefont {Penc}, \citenamefont {Troyer},\ and\ \citenamefont {Mila}}]{Corboz2011PRLSU4}%
  \BibitemOpen
  \bibfield  {author} {\bibinfo {author} {\bibfnamefont {P.}~\bibnamefont {Corboz}}, \bibinfo {author} {\bibfnamefont {A.~M.}\ \bibnamefont {L\"auchli}}, \bibinfo {author} {\bibfnamefont {K.}~\bibnamefont {Penc}}, \bibinfo {author} {\bibfnamefont {M.}~\bibnamefont {Troyer}},\ and\ \bibinfo {author} {\bibfnamefont {F.}~\bibnamefont {Mila}},\ }\bibfield  {title} {\bibinfo {title} {Simultaneous dimerization and su(4) symmetry breaking of 4-color fermions on the square lattice},\ }\href {https://doi.org/10.1103/PhysRevLett.107.215301} {\bibfield  {journal} {\bibinfo  {journal} {Phys. Rev. Lett.}\ }\textbf {\bibinfo {volume} {107}},\ \bibinfo {pages} {215301} (\bibinfo {year} {2011})}\BibitemShut {NoStop}%
\bibitem [{\citenamefont {Hasenfratz}\ and\ \citenamefont {Niedermayer}(1993)}]{Hasenfratz1993}%
  \BibitemOpen
  \bibfield  {author} {\bibinfo {author} {\bibfnamefont {P.}~\bibnamefont {Hasenfratz}}\ and\ \bibinfo {author} {\bibfnamefont {F.}~\bibnamefont {Niedermayer}},\ }\bibfield  {title} {\bibinfo {title} {Finite size and temperature effects in the af heisenberg model},\ }\href {https://doi.org/10.1007/BF01309171} {\bibfield  {journal} {\bibinfo  {journal} {Zeitschrift f{\"u}r Physik B Condensed Matter}\ }\textbf {\bibinfo {volume} {92}},\ \bibinfo {pages} {91} (\bibinfo {year} {1993})}\BibitemShut {NoStop}%
\bibitem [{\citenamefont {Ding}\ \emph {et~al.}(2018)\citenamefont {Ding}, \citenamefont {Zhang},\ and\ \citenamefont {Guo}}]{ding2018engineering}%
  \BibitemOpen
  \bibfield  {author} {\bibinfo {author} {\bibfnamefont {C.}~\bibnamefont {Ding}}, \bibinfo {author} {\bibfnamefont {L.}~\bibnamefont {Zhang}},\ and\ \bibinfo {author} {\bibfnamefont {W.}~\bibnamefont {Guo}},\ }\bibfield  {title} {\bibinfo {title} {Engineering surface critical behavior of ($2+1$)-dimensional o(3) quantum critical points},\ }\href {https://doi.org/10.1103/PhysRevLett.120.235701} {\bibfield  {journal} {\bibinfo  {journal} {Phys. Rev. Lett.}\ }\textbf {\bibinfo {volume} {120}},\ \bibinfo {pages} {235701} (\bibinfo {year} {2018})}\BibitemShut {NoStop}%
\bibitem [{\citenamefont {Rademaker}(2015)}]{rademaker2015tower}%
  \BibitemOpen
  \bibfield  {author} {\bibinfo {author} {\bibfnamefont {L.}~\bibnamefont {Rademaker}},\ }\bibfield  {title} {\bibinfo {title} {Tower of states and the entanglement spectrum in a coplanar antiferromagnet},\ }\href {https://doi.org/10.1103/PhysRevB.92.144419} {\bibfield  {journal} {\bibinfo  {journal} {Phys. Rev. B}\ }\textbf {\bibinfo {volume} {92}},\ \bibinfo {pages} {144419} (\bibinfo {year} {2015})}\BibitemShut {NoStop}%
\bibitem [{sm()}]{sm}%
  \BibitemOpen
  \href@noop {} {\bibinfo  {journal} {See Supplemental Material at http://link.aps.org/supplemental/10.1103/7j21-l3pg for detailed results obtained from DMRG and some analysis on TOS}\ }\BibitemShut {NoStop}%
\bibitem [{\citenamefont {Francesco}\ \emph {et~al.}(2012)\citenamefont {Francesco}, \citenamefont {Mathieu},\ and\ \citenamefont {S{\'e}n{\'e}chal}}]{francesco2012conformal}%
  \BibitemOpen
\bibfield  {journal} {  }\bibfield  {author} {\bibinfo {author} {\bibfnamefont {P.}~\bibnamefont {Francesco}}, \bibinfo {author} {\bibfnamefont {P.}~\bibnamefont {Mathieu}},\ and\ \bibinfo {author} {\bibfnamefont {D.}~\bibnamefont {S{\'e}n{\'e}chal}},\ }\href@noop {} {\emph {\bibinfo {title} {Conformal field theory}}}\ (\bibinfo  {publisher} {Springer Science \& Business Media},\ \bibinfo {year} {2012})\BibitemShut {NoStop}%
\bibitem [{\citenamefont {Zhu}\ \emph {et~al.}(2023)\citenamefont {Zhu}, \citenamefont {Han}, \citenamefont {Huffman}, \citenamefont {Hofmann},\ and\ \citenamefont {He}}]{zhu2023uncovering}%
  \BibitemOpen
  \bibfield  {author} {\bibinfo {author} {\bibfnamefont {W.}~\bibnamefont {Zhu}}, \bibinfo {author} {\bibfnamefont {C.}~\bibnamefont {Han}}, \bibinfo {author} {\bibfnamefont {E.}~\bibnamefont {Huffman}}, \bibinfo {author} {\bibfnamefont {J.~S.}\ \bibnamefont {Hofmann}},\ and\ \bibinfo {author} {\bibfnamefont {Y.-C.}\ \bibnamefont {He}},\ }\bibfield  {title} {\bibinfo {title} {Uncovering conformal symmetry in the 3d ising transition: State-operator correspondence from a quantum fuzzy sphere regularization},\ }\href {https://doi.org/10.1103/PhysRevX.13.021009} {\bibfield  {journal} {\bibinfo  {journal} {Phys. Rev. X}\ }\textbf {\bibinfo {volume} {13}},\ \bibinfo {pages} {021009} (\bibinfo {year} {2023})}\BibitemShut {NoStop}%
\bibitem [{\citenamefont {Sandvik}(2007{\natexlab{b}})}]{Sandvik2007}%
  \BibitemOpen
  \bibfield  {author} {\bibinfo {author} {\bibfnamefont {A.~W.}\ \bibnamefont {Sandvik}},\ }\bibfield  {title} {\bibinfo {title} {Evidence for deconfined quantum criticality in a two-dimensional heisenberg model with four-spin interactions},\ }\href {https://doi.org/10.1103/PhysRevLett.98.227202} {\bibfield  {journal} {\bibinfo  {journal} {Phys. Rev. Lett.}\ }\textbf {\bibinfo {volume} {98}},\ \bibinfo {pages} {227202} (\bibinfo {year} {2007}{\natexlab{b}})}\BibitemShut {NoStop}%
\bibitem [{\citenamefont {Wang}\ \emph {et~al.}(2022)\citenamefont {Wang}, \citenamefont {Ma}, \citenamefont {Cheng},\ and\ \citenamefont {Meng}}]{wangScaling2022}%
  \BibitemOpen
  \bibfield  {author} {\bibinfo {author} {\bibfnamefont {Y.-C.}\ \bibnamefont {Wang}}, \bibinfo {author} {\bibfnamefont {N.}~\bibnamefont {Ma}}, \bibinfo {author} {\bibfnamefont {M.}~\bibnamefont {Cheng}},\ and\ \bibinfo {author} {\bibfnamefont {Z.~Y.}\ \bibnamefont {Meng}},\ }\bibfield  {title} {\bibinfo {title} {{Scaling of the disorder operator at deconfined quantum criticality}},\ }\href {https://doi.org/10.21468/SciPostPhys.13.6.123} {\bibfield  {journal} {\bibinfo  {journal} {SciPost Phys.}\ }\textbf {\bibinfo {volume} {13}},\ \bibinfo {pages} {123} (\bibinfo {year} {2022})}\BibitemShut {NoStop}%
\bibitem [{\citenamefont {Wang}\ \emph {et~al.}(2025)\citenamefont {Wang}, \citenamefont {Deng}, \citenamefont {Liu}, \citenamefont {Wang}, \citenamefont {Ding}, \citenamefont {Zhang}, \citenamefont {Guo},\ and\ \citenamefont {Yan}}]{wang2024probing}%
  \BibitemOpen
  \bibfield  {author} {\bibinfo {author} {\bibfnamefont {Z.}~\bibnamefont {Wang}}, \bibinfo {author} {\bibfnamefont {Z.}~\bibnamefont {Deng}}, \bibinfo {author} {\bibfnamefont {Z.}~\bibnamefont {Liu}}, \bibinfo {author} {\bibfnamefont {Z.}~\bibnamefont {Wang}}, \bibinfo {author} {\bibfnamefont {Y.-M.}\ \bibnamefont {Ding}}, \bibinfo {author} {\bibfnamefont {L.}~\bibnamefont {Zhang}}, \bibinfo {author} {\bibfnamefont {W.}~\bibnamefont {Guo}},\ and\ \bibinfo {author} {\bibfnamefont {Z.}~\bibnamefont {Yan}},\ }\bibfield  {title} {\bibinfo {title} {Universal behavior in entanglement entropy reveals quantum criticality and underlying symmetry breaking},\ }\href {https://doi.org/10.1088/0256-307X/42/11/110712} {\bibfield  {journal} {\bibinfo  {journal} {Chinese Physics Letters}\ }\textbf {\bibinfo {volume} {42}},\ \bibinfo {pages} {110712} (\bibinfo {year} {2025})}\BibitemShut {NoStop}%
\bibitem [{\citenamefont {Yang}\ \emph {et~al.}(2023)\citenamefont {Yang}, \citenamefont {Vanhecke},\ and\ \citenamefont {Schuch}}]{yang2023detecting}%
  \BibitemOpen
  \bibfield  {author} {\bibinfo {author} {\bibfnamefont {M.}~\bibnamefont {Yang}}, \bibinfo {author} {\bibfnamefont {B.}~\bibnamefont {Vanhecke}},\ and\ \bibinfo {author} {\bibfnamefont {N.}~\bibnamefont {Schuch}},\ }\bibfield  {title} {\bibinfo {title} {Detecting emergent continuous symmetries at quantum criticality},\ }\href {https://doi.org/10.1103/PhysRevLett.131.036505} {\bibfield  {journal} {\bibinfo  {journal} {Phys. Rev. Lett.}\ }\textbf {\bibinfo {volume} {131}},\ \bibinfo {pages} {036505} (\bibinfo {year} {2023})}\BibitemShut {NoStop}%
\bibitem [{\citenamefont {Mao}\ \emph {et~al.}(2026)\citenamefont {Mao}, \citenamefont {Wang}, \citenamefont {Chen},\ and\ \citenamefont {Yan}}]{dataset}%
  \BibitemOpen
  \bibfield  {author} {\bibinfo {author} {\bibfnamefont {B.-B.}\ \bibnamefont {Mao}}, \bibinfo {author} {\bibfnamefont {Z.}~\bibnamefont {Wang}}, \bibinfo {author} {\bibfnamefont {B.-B.}\ \bibnamefont {Chen}},\ and\ \bibinfo {author} {\bibfnamefont {Z.}~\bibnamefont {Yan}},\ }\href@noop {} {\bibfield  {journal} {\bibinfo  {journal} {Zenodo, https://doi.org/10.5281/zenodo.18206451}\ } (\bibinfo {year} {2026})}\BibitemShut {NoStop}%
\end{thebibliography}%

\newpage 
\onecolumngrid
\cleardoublepage
\setcounter{section}{0}
\setcounter{page}{1}
\setcounter{equation}{0}
\setcounter{figure}{0}
\renewcommand{\theequation}{S\arabic{equation}}
\renewcommand{\thefigure}{S\arabic{figure}}
\renewcommand{\thetable}{S\arabic{table}}

\thispagestyle{empty}
\begin{widetext}

\begin{center}
{\large{\bfseries{Supplement Materials for}}}

{\large{\bfseries{Detecting the (emergent) continuous symmetry of criticality via a subsystem's entanglement spectrum}}}
  
  \vspace{1em}

Bin-Bin Mao,$^{1, 2, *}$ Zhe Wang,$^{1, 3, *}$ Bin-Bin Chen,$^{4, \dagger}$ and Zheng Yan$^{1, 3, \ddagger}$
\vspace{1em}

$^1$\textit{Department of Physics, School of Science and Research Center for Industries of the Future, Westlake University, Hangzhou 310030, China}

$^2$\textit{School of Foundational Education, University of Health and Rehabilitation Sciences, Qingdao 266000, China}

$^3$\textit{Institute of Natural Sciences, Westlake Institute for Advanced Study, Hangzhou 310024, China}

$^4$\textit{Peng Huanwu Collaborative Center for Research and Education, Beihang University, Beijing 100191, China}
\vspace{2ex}
\end{center}

\vskip3em

\section{{TOS via DMRG simulations}}
\label{sec:I}
In Supplemental Materials \ref{sec:I}, we show more detailed data of the tower of states (TOS) obtained from density matrix renormalization group (DMRG).

\subsection{More on the dimerized Heisenberg model}
In principle, the idea of detecting symmetry by TOS should also work for tensor-network-based methods. As shown in Fig.\ref{figs1}, we have included the DMRG results to probe the emergent $O(3)$ symmetry at a (2+1)d $O(3)$ critical point of a dimerized Heisenberg model. 
Specifically, we have performed DMRG simulations on Y cylinder (YC) $L_y\times L_x$ geometry with `Y' indicating periodic boundary condition along $y$ direction and open boundary condition along $x$ direction [c.f. Fig.\ref{figs1}~(a)]. The DMRG simulation is then carried out along the so-called snake-like path, i.e., the 1D index $i$ (ranging from 0 to $L_xL_y-1$) of the local tensor of the matrix product state/operator (MPS/O) is chosen as $i=(x\times L_y + y)$ with $(x,y)$ being the 2D coordinate of the corresponding site in the lattice. The bond dimension is kept up to 1024 SU(2)-rotation-invariant multiplets (equivalently $\sim 4000$ U(1) states) to ensure small truncation errors with $O(10^{-9})$ for YC$6\times24$ simulations and $O(10^{-6})$ for YC$8\times16$ simulations.

\begin{figure}[htp]
\centering
\includegraphics[width=0.8\textwidth]{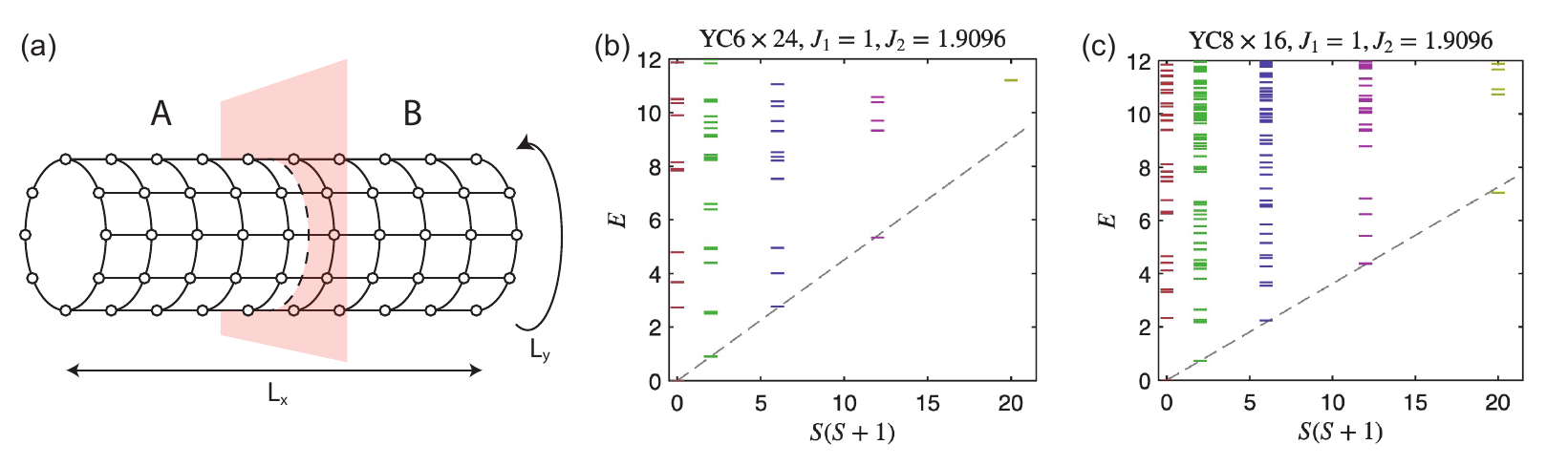}
\caption{(a) Schematic plot for the bipartition-cut used in DMRG simulations.
(b) For YC6$\times$24 cylinder, the entanglement spectrum is plotted versus $S(S+1)$, exhibiting low-lying Anderson TOS. (c) For YC8$\times$16 cylinder, the entanglement spectrum is plotted versus $S(S+1)$, exhibiting low-lying Anderson TOS.}
\label{figs1}
\end{figure}

We have to emphasize that the subsystem $A$ is not limited to a 1D chain, it also works for a bipartition cut which separate the cylinder into $A$ and $B$ sub-cylinder as shown in Fig.\ref{figs1}~(a). According to the properties of different numerical methods, we can choose the suitable cut. In quantum Monte Carlo (QMC), 1D subsystem is easy for simulation while half-cut is better for DMRG, thus we show the DMRG results under half-cut without corner. As mentioned in the main text, TOS of entanglement spectrum only works in cornerless cut.

It can be seen that, when tuning $J_2/J_1$ to the $O(3)$ transition point, the low-lying levels in the obtained entanglement spectrum follow a nice linear scaling behavior versus $S(S+1)$, exhibiting the Anderson Tower of States $O(3)$. 
This showcases the generality of our idea of probing the emergent symmetries via the entanglement TOS [Fig.1 of the main text, the phase diagram of entanglement Hamiltonian], which is not merely restricted to QMC, but also works for tensor network methods.

\end{widetext}

\section{Determination of $N$ and finite size effect of TOS}

In this section, we give the process of determining $N$ and the finite size effect of TOS. One can find out that the properties of TOS are perfectly preserved at finite size, thus it is possible to assess spontaneous symmetry breaking (SSB) intuitively with limited computational resource. We can determine the symmetry of a system at its critical point in systems of small size.
\begin{figure}[h!]
    \centering
\begin{minipage}[c]{0.25\textwidth}
    \includegraphics[width=\textwidth]{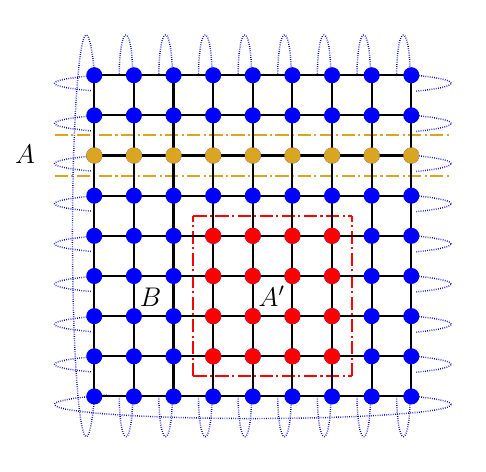}
    \end{minipage}

\caption{The geometry of different subsystem on the square lattice. 
The dashed lines are used to illustrate the bipartition into two subsystems. The yellow dashed lines illustrate the cutting method that $A$ is a ring and the red dashed lines illustrate the method that $A^\prime$ is a block. For each cutting method, the other part is denoted as $B$. }
	\label{fig:lattice}
\end{figure}

For a $d\geq 2$ dimensional quantum $O(N)$ model, its ground state spontaneously beaks the continuous symmetry and is labeled by N\'eel order. Meanwhile, its energy spectrum has the characteristic of tower of states (TOS) structure. The energies of the TOS in finite size are
\begin{equation}
E_S(L)-E_0(L)=\frac{S(S+N-2)}{2\chi_{\perp}L^d}
\label{eq3}
\end{equation}
%$E_S(L)-E_0(L)=S(S+N-2)/(2\chi_{\perp}L^d)$~\cite{hasenfratz1993finite,Deng2023improved}, 
where $S$ is the total spin momentum of the system, $\chi_{\perp}$ is the transverse susceptibility in the thermodynamic limit. Except the energy spectrum, the ES is also expected having TOS when the ground state is continuous symmetry breaking. 
The TOS of entanglement spectrum requires that the entangled region has no corners. 

In the following, we show the results of $J-Q_3$ model with two different cut-geometries as shown in Fig. ~\ref{fig:lattice}.
One subsystem chooses the $A$ as a chain without corner as the yellow dots ($A$ region) in Fig.~\ref{fig:lattice} displays, the other chooses $A$ as a block region with four corners as the red dots ($A'$ region) in Fig.~\ref{fig:lattice}. We will demonstrate the TOS only works well in cornerless cut.

In the calculation, we choose the system as $L_x=Ly=8$, $\beta=100$ and the number of samplings is $10^{10}$. To quantify how well the TOS structure exhibits a linear relationship, we compute the root-mean-square error (RMSE)
\begin{equation}
    \mathrm{RMSE} = \sqrt{\frac{1}{n}\sum_{i}^{n}(x_i-x_i^\prime)^2}
\end{equation}
where $x_i^\prime$ is the predicted value.

Since the importance sampling makes that the low-lying levels converge first, we can obtain a predicted linear-fit-line using the two lowest levels.  Then we can compute the RMSE between the TOS data and the fitted data.  Since $N$ corresponds to $O(N )$ symmetry, our calculations restrict $N$ to be an integer.
\begin{figure}[h!]
    \centering
\begin{minipage}[c]{0.4\textwidth}
    \includegraphics[width=\textwidth]{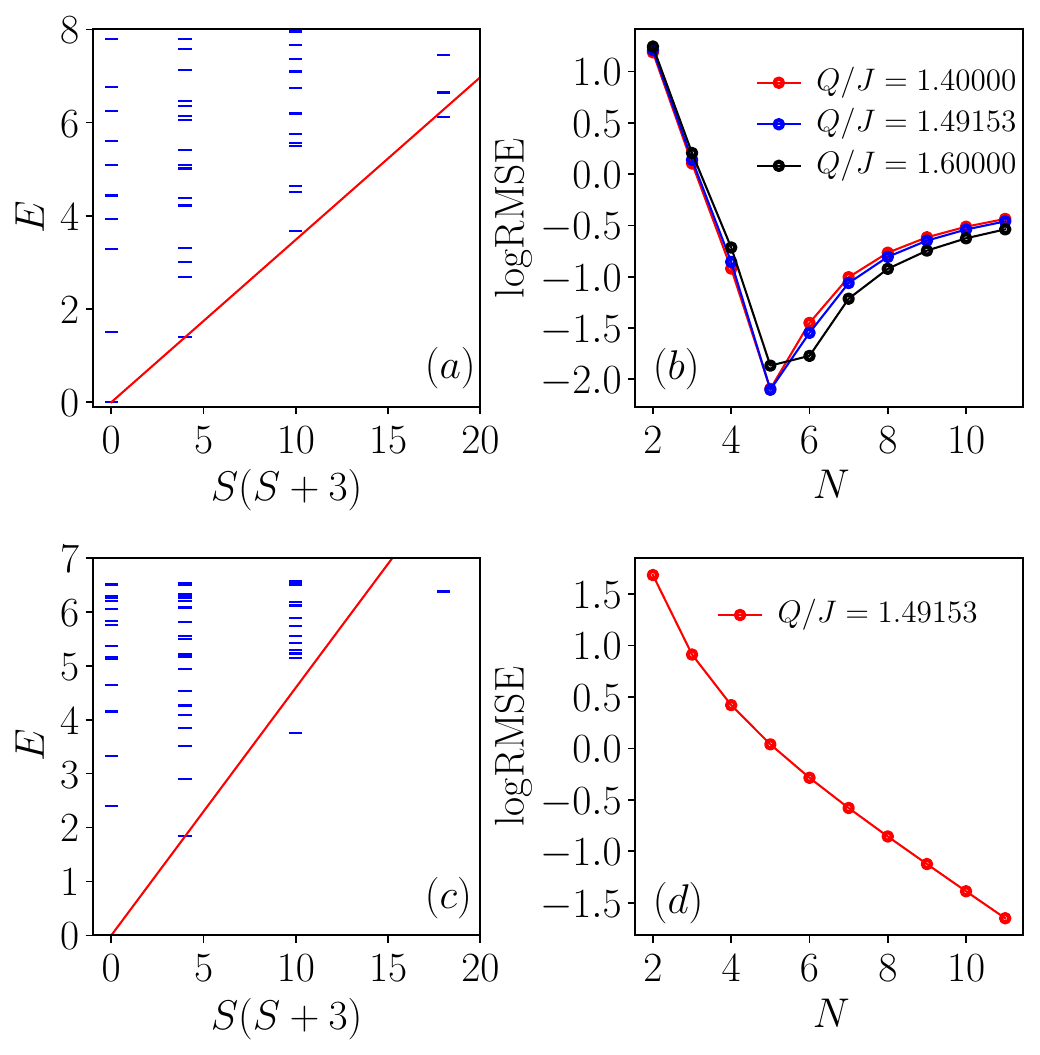}
    \end{minipage}

\caption{Entanglement spectrum (ES) of the $J-Q_3$ square lattice model with system size $L_x=L_y=8$ and the Root Mean Square Error(RMSE) of linear TOS structure.(a) ES for the subsystem chosen as a chain. (b) RMSE of TOS for the subsystem chosen as a chain with various $Q$. (c)ES for the subsystem chosen as a block with size $4\times 4$. (d) RMSE of TOS for the subsystem chosen as a block with size $4\times 4$.  All the error bars are smaller than the data symbols.}
	\label{fig:jq3}
\end{figure}

\begin{figure*}[htp]
    \centering
    \begin{minipage}[c]{0.85\textwidth}
    \includegraphics[width=\textwidth]{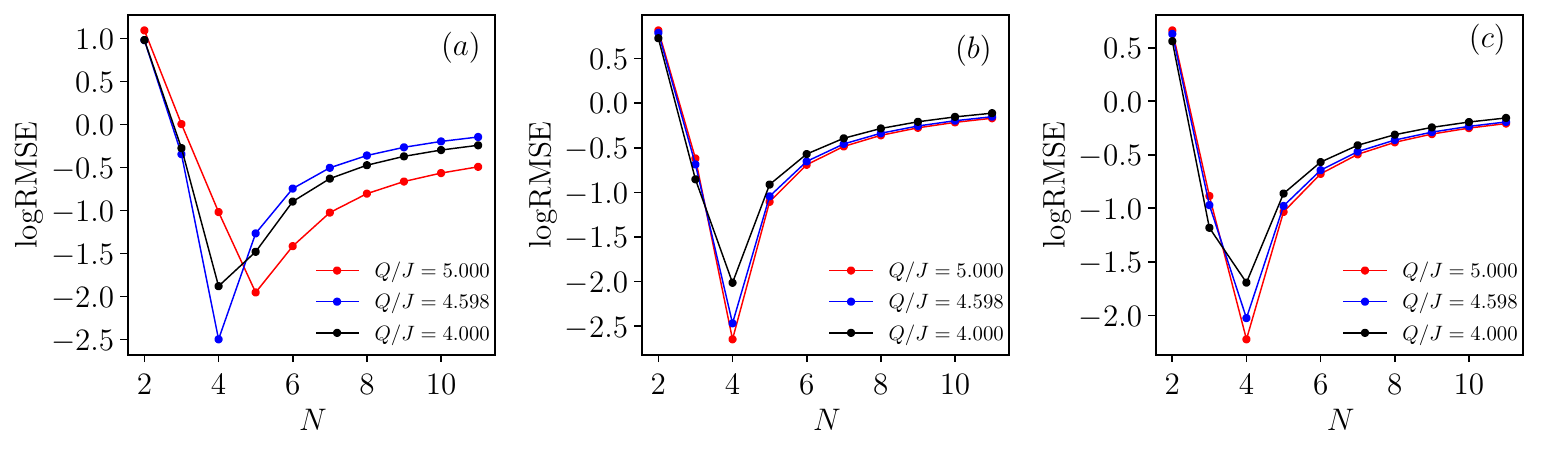}
    \end{minipage}
\caption{Root mean square error(RMSE) of checkerboard $J-Q$ model. (a) $L_x =L_y = 8$. (b) $L_x =L_y = 10$. (c) $L_x =L_y = 12$.  All the error bars are smaller than the data symbols.}
	\label{fig:cbjq}
\end{figure*}

Fig.~\ref{fig:jq3} (a) shows ES of the $J-Q_3$ model at the critical point that a chain is chosen as the subsystem. ES shows a well linear relation with $N=5$ which reveals the emergent $SO(5)$ symmetry at the deconfined quantum critical point (DQCP). Fig.~\ref{fig:jq3} (b) shows RMSE as a function of $N$ with various $Q$. It is obvious that RMSE has a smallest value at $N=5$, it means that $N=5$ gives the best linear fitting relation. For $Q=1.4$ and $Q=1.6$, RMSE also has the smallest value at $N=5$, this indicates that the TOS structure holds well near the critical point region. Thus, TOS is a robust character near the critical point and can be used to determine the emergent symmetry.

For the condition that a block is chosen as the subsystem, Fig.~\ref{fig:jq3} (c) shows that the high-level points deviate significantly from the fitting line. Here, if we forcibly choose $N=5$ under the conered cut. It supports the subsystem should be cornerless again.
Fig.~\ref{fig:jq3} (d) shows RMSE as a function of $N$, one can find out that RMSE decreases with the increase of $N$ and has no deep minimum. It indicates that the linear function can not give the best $N$. The linear TOS structure con not hold in this corner subsystem.

In the following, we check the finite size effect of TOS structure by the checkerboard $J-Q$ model with emergent $O(4)$ symmetry. We calculated RMSE of ES for various system size $L=8, 10, 12$ and $Q/J=4, 4.598$ (phase transition point) and $5$, the results are shown in Fig. \ref{fig:cbjq}. The subfigure (a) with system size $L_x =L_y = 8$ indicates that the ES shows a well linear TOS structure at $Q/J=4$ and $Q/J=4.598$. For larger size system as shown in Fig.~\ref{fig:cbjq}(b) and (c), we can find out that ES shows a well linear TOS structure in a region near the phase transition point. All of them can give the same $N=4$ at the critical point, indicating the system has $O(4)$ spontaneous symmetry breaking. The structure of TOS and the value of $N$ has no finite size effect at the phase transition point. Thus, we can determine the symmetry at phase transition point by the data of the system with small size, e.g. $L_x=Ly=8$

\end{document}